\documentclass[english]{article}
\usepackage{custom_tex}
\usepackage[table]{xcolor}
\usepackage{setspace}
\usepackage{geometry}
\usetikzlibrary{positioning,calc,arrows.meta}

\title{Statistical inference in two-stage observation models including algorithmic randomness}

\author[1]{Zhixiang Zhang}
\author[2]{Sokbae Lee}
\author[3]{Edgar Dobriban\footnote{Author e-mail addresses: 
\texttt{zhixzhang@um.edu.mo},
\texttt{sl3841@columbia.edu},
\texttt{dobriban@wharton.upenn.edu}
}}

\affil[1]{Department of Mathematics, 
University of Macau}

\affil[2]{Department of Economics,
Columbia University}

\affil[3]{Department of Statistics and Data Science,  
University of Pennsylvania}

\begin{document}

\maketitle

\begin{abstract}
Randomized algorithms, such as random sampling, random projections, and
stochastic optimization, are increasingly used to reduce the computational
cost of modern statistical analysis. These algorithms introduce algorithmic
randomness in addition to the sampling randomness in the data, and this extra
source of variation complicates statistical inference. We develop a framework
for valid inference in such two-stage observation models, where data are first
generated from an underlying population process and are then analyzed through
a randomized algorithm.

Our method, called sub-randomization, runs the randomized algorithm multiple
times at different computational scales and uses the auxiliary runs to
approximate the conditional algorithmic error distribution. In important
converging-scale settings, the procedure avoids estimating the limiting
covariance matrix or other nuisance parameters in the limiting law. We
illustrate the method in two settings where standard approaches can fail to
achieve nominal coverage: inference from repeated observations with highly
correlated noise, and confidence sets for the minimizers of stochastic
optimization problems computed using momentum methods, with particular
emphasis on the stochastic heavy ball algorithm.
\end{abstract}

\tableofcontents

\medskip

\section{Introduction}

Since large and complex datasets are common across scientific and applied domains, there is increasing demand for data analysis methods that are both statistically reliable and computationally fast. Randomized algorithms are now widely used for this purpose, including Monte Carlo methods \citep[e.g.,][]{mcbook,practicalqmc}, random sampling and random projections \citep[e.g.,][]{vempala2005random, mahoney2011randomized,woodruff2014sketching,Lee:Ng:2020}, randomized methods for principal component analysis \citep[e.g.,][]{halko2011algorithm}, as well as stochastic optimization \citep[e.g.,][]{spall2005introduction,bottou2018optimization}. These methods can substantially reduce computational cost, but they also introduce algorithmic randomness, so repeated runs on the same data may give different outputs. 
Therefore, in addition to the sampling variability in the data set, there is also variability in the algorithmic output. 
This raises a basic inferential question: how should we perform statistical inference when the randomness arises from both algorithmic and sampling variation?

As an empirical 
illustration, consider
the IPUMS dataset \citep{ruggles2022ipums}, which aggregates U.S. census microdata and is therefore large, with millions of observations.
Fitting regression models for earnings can be computationally challenging, especially when the analysis is repeated many times, as in exploratory workflows where multiple specifications are fit.

To address this computational bottleneck, one may turn to randomized algorithms, such as running the least squares estimate on a random subset of the data. 
Using such algorithms accelerates computation, but this comes at the cost of additional algorithmic variation. 
Thus, uncertainty arises not only from the data generating process---which may exhibit cluster or spatial correlation that is already difficult to model---but also from algorithmic randomness. This motivates the need for inference methods 
that can utilize computationally cheaper randomized algorithms while still providing valid statistical coverage.

\begin{figure}
    \centering
    \includegraphics[width=0.7\linewidth]{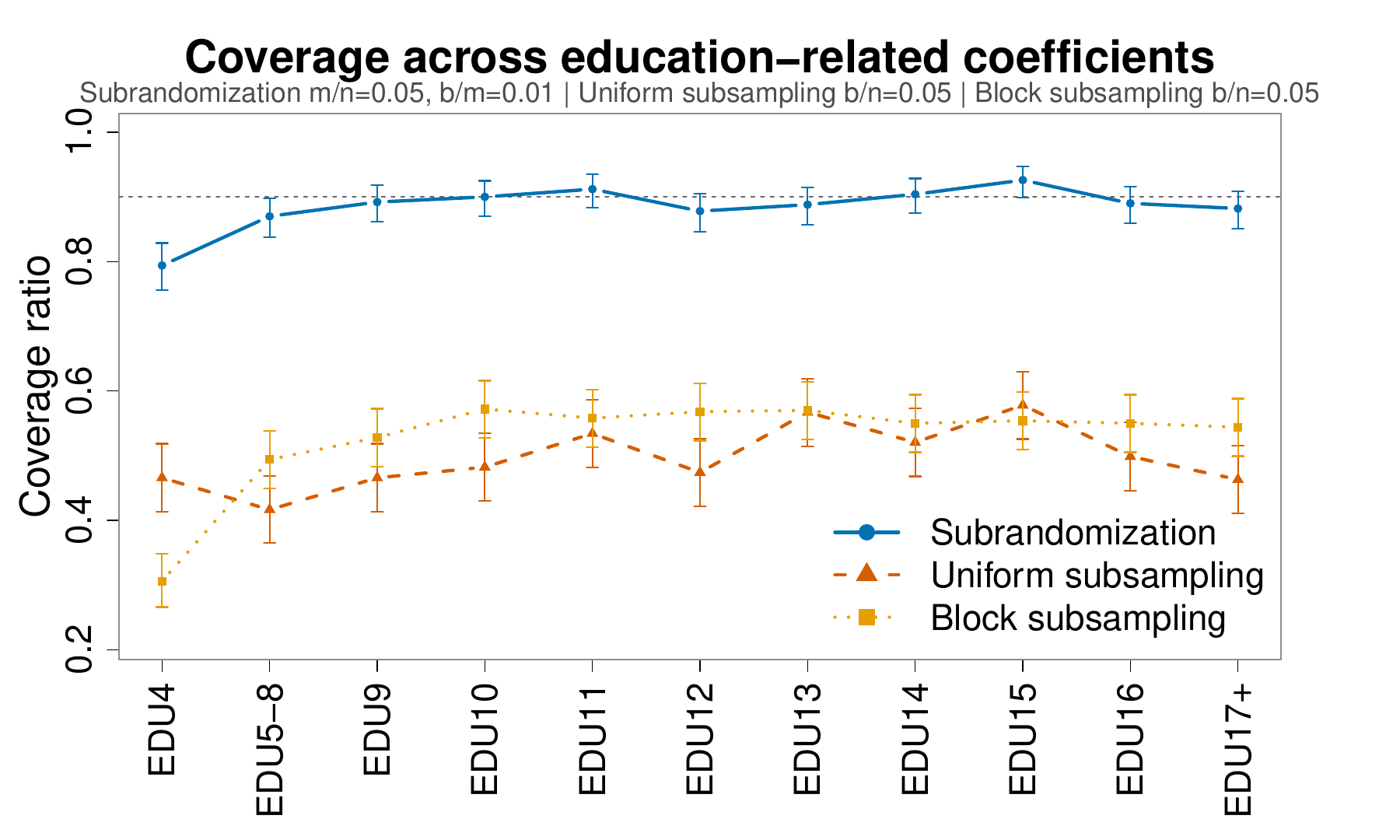}
    \caption{\footnotesize 
    Results on the IPUMS dataset, showing empirical coverage probabilities for the coefficients of regressing log hourly wage on several education-level covariates (EDU4 through EDU17+), with noise that is correlated across individuals within clusters corresponding to the same education level.
    We observe that our proposed sub-randomization method achieves the desired coverage level while methods based on uniform and block subsampling achieve significantly lower coverage.}
    \label{fig:edu_coverage}
\end{figure}

Work over the last decade 
\citep[e.g.,][]{lopes2018error,ahfock2021statistical,lee2022least,ma2022asymptotic}
has studied statistical inference problems in the important example of least squares regression with sampling or sketching algorithms, 
as well as for certain specific stochastic optimization algorithms \citep[e.g.,][]{toulis2017asymptotic,fang2018bootstrap,chen2020statistical,zhu2023online}.
However, existing inferential guarantees are typically tied to particular
randomized algorithms, particular estimators, or settings in which the
unknown quantities in the limiting law can be consistently estimated. A
broadly applicable framework for inference from randomized algorithmic
outputs remains less developed.

This is precisely the methodology we aim to develop here.  
We study statistical inference in a two-stage setting where data are generated from an underlying population model, but
data analysis involves randomized algorithms. In the population stage, the data $\mD_n$ are produced by an unknown data-generating process associated with population parameter $\theta^\star$. 
However, direct full-data computation on $\mD_n$ may be infeasible or
undesirable, typically because of its size or because repeated full-data
computations are too costly. We instead assume access to randomized
algorithms that query or process $\mD_n$ at a prescribed computational scale,
yielding an estimator $\widehat\theta_m$ of the population parameter or of the
full-data target.  
This perspective covers, among others, 
random sampling and projection methods, 
and
streaming stochastic optimization methods such as stochastic gradient descent (SGD).

In this paper, we develop \emph{sub-randomization}, an inference method based
on repeated runs of a randomized algorithm at different computational scales.
The method is inspired by subsampling
\citep{politis1994large,politis1999subsampling}, but it targets a different
law: the conditional distribution of the randomized algorithmic output around
the full-data target. In many settings this law can be approximated without
estimating the limiting covariance matrix or other nuisance parameters.

We first illustrate sub-randomization in least-squares regression with a
general noise covariance structure. The method is robust to forms of
dependence that can make ordinary subsampling fail to attain nominal coverage.
See \Cref{fig:edu_coverage} for an example using the IPUMS dataset where both ordinary and block subsampling fail to achieve nominal coverage while our proposed method succeeds.
We then consider stochastic optimization, focusing on momentum-based methods.
While much of the existing inferential literature for stochastic optimization
is built around SGD or averaged SGD, practitioners often use accelerated or
momentum-based procedures such as stochastic heavy ball. We show how
sub-randomization can provide inference for such outputs while avoiding online
covariance estimation.

The remainder of the paper is organized as follows. Section~\ref{sec:two_stage} introduces the two-stage model. Section~\ref{sec:subrand} presents the sub-randomization method, establishes its inferential validity, 
and relates it to classical subsampling. 
Section~\ref{sec:examples} develops examples including least squares with 
general correlation structures in the noise,
as well as stochastic optimization methods such as stochastic heavy ball.
The Supplementary Material contains the proofs and supplementary numerical simulations.

\subsection{Related work}
\label{relw}

There is a large literature on randomized algorithms, including stochastic approximation and stochastic optimization; see, for example,
\cite{kushner2003stochastic,borkar2009stochastic,benveniste2012adaptive,bottou2018optimization,powell2019unified},
as well as on sketching and random projection methods \cite{vempala2005random, li2006very, halko2011finding, mahoney2011randomized, woodruff2014sketching, drineas2016randnla,martinsson2020randomized}.

Existing statistical inference approaches based on randomized algorithmic outputs can be grouped into two broad classes. The first class is pivotal. These methods derive an asymptotic distribution for the randomized output and then estimate the unknown scaling quantities in the limit law, most often an asymptotic covariance matrix in Gaussian limits. In randomized least squares, for example, inference can be based on limit theory for sketching or sampling estimators together with consistent covariance estimation \citep{ahfock2021statistical,lee2022least,ma2022asymptotic}. In stochastic optimization, a large literature develops online methods for estimating the asymptotic covariance matrix of averaged SGD and related stochastic approximation procedures \citep{chen2020statistical,zhu2023online,roy2023online}. These approaches are powerful when  the unknown parameters in the limiting laws can be reliably estimated. 
Another line of research avoids online covariance estimation by constructing self-normalized, asymptotically pivotal statistics for stochastic optimization. The random scaling method of \cite{lee2022fast} is a prominent example, with extensions developed for Markovian data \citep{li2023online}, heavy-tailed noise \citep{li2024stochastic}, and constrained optimization settings \citep{du2025online}.

A second class uses resampling-based inference, represented by bootstrap methods. \citet{lopes2018error,lopes2020error} approximate algorithmic error by drawing bootstrap samples from quantities produced by a randomized algorithm.  \citet{lopes2019estimating} use bootstrap ideas to estimate the variance of bagging and random forests. 
Resampling methods are attractive because they can avoid explicit analytic covariance calculations, but their validity must account for two distinct randomization procedures: the original randomized algorithm and the second-stage resampling scheme.
In contrast, our approach requires only the limiting law of the original randomized procedure at two computational scales.

Our work is most closely related to subsampling. 
Subsampling \citep{politis1994large,politis1999subsampling} and the closely related delete-$d$ jackknife \citep{wu1986jackknife,shao1989general} 
and $b$-out-of-$n$ bootstrap 
\citep{bickel2008choice}
have been widely studied and developed;
see also \citet{politis2023scalable} for aggregation of deterministic subsamples.
Subsampling is designed to approximate the sampling distribution of an estimator $\theta_n$ viewed as an estimator of the population parameter $\theta^\star$, 
whereas sub-randomization aims to estimate the conditional distribution of algorithmic error for a randomized estimator $\htheta_m$, 
centered at $\theta_n$. 
For this reason, since sub-randomization effectively views
$\theta_n$ as a ``target", 
we will refer to $\theta_n$ as the full-data target.

{\bf Notation.}
For a positive integer $d\ge 1$, we denote $[d] = \{1,\ldots,d\}$.
We write $\Spp$ for the set of $d \times d$ 
positive definite matrices.  Denote the $i$-th largest eigenvalue of a Hermitian matrix $H$ by $\lambda_{i}(H)$. 
We use $O(\cdot)$ and $o(\cdot)$ for the standard big-O and little-o notation. 
For a sequence  $(a_n)_{n\ge 1}$ of scalars, 
we write $a_n = O_P(1)$ if $(a_n)_{n\ge 1}$ is bounded in probability and $a_n = o_P(1)$ if $(a_n)_{n\ge 1}$ converges to zero in probability.
For any vector $a$ or matrix $A$, we let $\|a\|$ and $\|A\|$ denote their Euclidean and spectral norms, respectively. 
For a sequence  $(a_n)_{n\ge 1}$ of vectors  or $(A_n)_{n\ge 1}$ of matrices with fixed dimension, we write $a_n = O_P(1)$ if $\|a_n\| =O_P(1)$; and $A_n=O_P(1)$ if $\|A_n\|=O_P(1)$.
The law of a random variable $X$ will be denoted by $\mathcal{L}(X)$; in displayed formulas we also use the shorthand $\Law(X)$. 
For random probability measures $\mu_n$, we write $\mu_n\Rightarrow_P \mu$ to mean weak convergence in probability, equivalently $\int f\,d\mu_n\to_P \int f\,d\mu$ for every bounded continuous function $f$. 
We write $a\odot b$ for elementwise multiplication, $1_n$ for the $n$-vector of ones, and $A\preceq B$ for the Loewner order on symmetric matrices.

\section{A two-stage model: population and algorithmic stage}\label{sec:two_stage}

\begin{figure}[t]
\centering
\begin{tikzpicture}[
  >=Latex,
  every node/.style={
    draw, rectangle,
    align=center,
    minimum width=2.35cm,
    minimum height=1.65cm
  },
  arr/.style={->, thick, shorten <=4pt, shorten >=4pt},
  scale=0.8, transform shape
]
\node (theta) {Population\\parameter\\$\theta^\star$};

\node (data) [right=1.55cm of theta] {Full dataset\\$\mD_n\sim P_{\theta^\star}$\\(limited access)};

\node (obs)  [right=1.75cm of data, text width=4.15cm] {Observed output\\$Z_{m,n}=\mA_m(\mD_n,S_{m,n})$\\(algorithm $\mA_m$)};

\node (gen)  [below=1.25cm of data, text width=4.15cm] {Auxiliary outputs\\$Z_{b,n,i}=\mA_b(\mD_n,S_{b,n,i})$\\$i=0,1,\dots,K_{m,n}$\\(algorithm $\mA_b$ may differ)};

\node (L)    [right=1.75cm of gen, text width=3.55cm] {Estimated error\\distribution\\$L_{b,m,n}$};

\node (C)    [right=1.75cm of L, text width=4.15cm] {Confidence set\\for $\theta_n(\mD_n)$\\and, under negligible sampling error,\\for $\theta^\star$};

\draw[arr] (theta) -- node[above, draw=none] {$P_{\theta^\star}$} (data);
\draw[arr] (data) -- (obs);
\draw[arr] (data) -- (gen);
\draw[arr] (gen) -- (L);
\draw[arr] (L) -- (C);
\draw[arr] (obs) -- (C.north);

\end{tikzpicture}
\caption{\footnotesize Our setting: population observation stage followed by algorithmic stage.}
\label{fig:framework}
\end{figure}

We now introduce the observation model
 considered in this paper; see
Figure~\ref{fig:framework} for a visual representation. 
There are two stages: 
a population sampling stage, 
and an algorithmic  stage. 

{\bf Population stage.}
The full dataset $\mD_n$, $n\ge 1$, follows an 
unknown distribution $P_{\theta^\star}$, where 
$\theta^\star\in\Theta\subseteq\R^d$, $d\ge 1$, is 
the true unknown population parameter.
Our goal is to construct a confidence set for $\theta^\star$.
 We focus on the setting 
 where only certain limited forms of computation and analysis can be performed on
 $\mD_n$,
 because $\mD_n$ is too large and computation is too expensive. 
 As a concrete example, this means that one might not be able to compute least squares regression on the full dataset, but only on a randomly projected dataset.
 The types of computation that can be run depend on the setting. We allow a broad range of algorithms, only requiring that their outputs satisfy certain distributional properties, as discussed later.

 As an intermediate part of the analysis, 
 it will also be helpful to
 consider inference for an auxiliary \emph{finite-sample target} $\theta_n=\theta_n(\mD_n)$, which we define on a case-by-case basis.
 Often, $\theta_n$ is the estimator that we would have liked to 
 use,
 if we had unlimited computational access to the full dataset $\mD_n$.
 For instance, in regression problems, this can be the least squares regression estimator on the full data set $\mD_n$. 
 However, since we do not have full access to $\mD_n$, we cannot actually compute $\theta_n$, 
 but instead can only use it as a theoretical device in our analysis.
 The reason why considering $\theta_n$ in the analysis is useful is that for 
 our large-data regime, 
 $\theta_n$ is often sufficiently close to $\theta^\star$
 that 
inferential statements for $\theta_n$ will transfer to 
$\theta^\star$. 
We will make this rigorous in what follows.

{\bf Algorithmic observation stage.}
Next, we consider the algorithmic observation stage,
which specifies how the dataset $\mD_n$ can be accessed. 
Since the dataset $\mD_n$ is too large to be accessed directly,
 we instead assume only that 
we can work with the output of a 
 randomized algorithm applied to $\mD_n$, 
 such as 
randomly subsampling $m$ out of $n$ datapoints  
or applying a random projection to $\mD_n$.
 Randomness is crucial, because we will use the 
 distribution of the algorithmic output as a basis for statistical inference.

To formalize this  algorithmic observation model,
we denote 
the algorithmic output by
$Z_{m,n}=\mA_m(\mD_n,S_{m,n})$,
where $\mA_m:\mathsf D_n\times\mathsf S_{m,n}\to\mathsf Z_{m,n}$ is a user-chosen randomized algorithm  and $S_{m,n}$ is a user-controlled source of randomness (e.g., a random sequence of indices), independent of $\mD_n$. 
We assume that we have full computational access to 
the algorithmic output
$Z_{m,n}$, 
and can compute desirable estimators using it as an input. 
Specifically, using $Z_{m,n}$, we compute an estimator $\htheta_m=\htheta_m(Z_{m,n})\in\R^d$ of $\theta^\star$; such as least squares regression on the subsampled data.

In addition,
 in order to 
perform statistical inference, our approach is to run the randomized algorithm on a smaller scale multiple times, 
leveraging the variability of these outputs to estimate the conditional algorithmic error distribution of $\htheta_m-\theta_n$ and, when the sampling error is negligible, to form confidence sets for $\theta^\star$.
 This approach applies when 
the variability of the randomized estimators coincides after proper rescaling over multiple scales.
This holds, for instance, when the estimators satisfy a central limit theorem, but also holds more broadly, as we will see in our examples. 
 
Specifically, 
for a smaller index $b<m$ and an integer $K_{m,n}\ge1$, we 
generate $K_{m,n}+1$ additional outputs via
a randomized algorithm
$Z_{b,n,i}=\mA_b(\mD_n,S_{b,n,i})$,
for $i=0,1,\dots,K_{m,n}$,
where $S_{b,n,i}$ are i.i.d.\ and independent of $(\mD_n,S_{m,n})$. 
For instance, this could mean taking smaller random samples of size $b$ from $\mD_n$.
While we allow $Z_{b,n,i}$ to be obtained via an algorithm different from that for $Z_{m,n}$, it will often be the same algorithm with a different scale, for instance running a stochastic optimization procedure for fewer steps. 
Using each $Z_{b,n,i}$, we compute $\htheta_{b,i}=\htheta_b(Z_{b,n,i})\in\R^d$, 
for instance a least squares estimator based on the smaller subsample.
 We will later explain how to use these estimators to perform inference on $\theta^\star$.

The framework includes 
a number of special cases. 
We now explain our notation in the context of some 
warmup problems.

\begin{example}\label{ex1}
(Sampling for least squares).
Consider an $n\times p$ design matrix $\mathbf{X}_n$ and a response vector $y_n\in\R^n$.
The full data are
\(
\mD_n=(\mathbf{X}_n,y_n),
\)
but the dataset $(\mathbf{X}_n,y_n)$ 
is too large to analyze directly.
We assume the population-level
linear regression model
\begin{equation}\label{lm_model}
y_n=\mathbf{X}_n\theta^\star+u_n,
\end{equation}
where $u_n$ is zero-mean random noise.
 We do not assume that $u_n$ has i.i.d.~coordinates; 
 in fact a key strength of our approach is that it can handle correlated noise. 
 
Our goal is to form a confidence region for the 
population regression parameter
\(
\theta^\star;
\)
or possibly a function (such as a coordinate) of this parameter.
If $\mathbf{X}_n$ has full column rank, the full-sample least squares estimator is
\(
\theta_n:=\argmin_{\theta\in\R^p}\|y_n-\mathbf{X}_n\theta\|_2^2=(\mathbf{X}_n^\top \mathbf{X}_n)^{-1}\mathbf{X}_n^\top y_n,
\)
which we consider as the finite-sample target $\theta_n$ of interest.
 However, this estimator is too expensive to compute.

Instead we consider a smaller sample size $m\ll n$, and we compute least squares on randomly sampled subsets of size $m$.
Let $I_m \subset \{1,\dots,n\}$ denote a uniformly sampled index set of size $m$, and let $\mathbf{X}_{n,I_m}$ and $y_{n,I_m}$ be the corresponding sampled rows of $\mathbf{X}_n$ and $y_n$. When $\mathbf{X}_{n,I_m}^\top\mathbf{X}_{n,I_m}$ is invertible, the uniform
subsampling least squares estimator is
\(
\widehat\theta_m
= \argmin_{\theta\in\R^p}\|y_{n,I_m} - \mathbf{X}_{n,I_m}\theta\|_2^2.
\)
On the exceptional event where the sampled Gram matrix is singular, one may use the Moore--Penrose pseudoinverse or define the estimator arbitrarily; under the regularity conditions used later, this event has probability tending to zero.
 In order to perform statistical inference,
 we also consider an even smaller sample size $b\le m$.  
We construct auxiliary least squares
estimators
on 
independently subsampled datasets of size $b$.

\end{example}

\begin{example}\label{example:so_general}
(Stochastic optimization).
Suppose that observations $\zeta_i= (X_i,Y_i)$, $i=0,1,\ldots$, are i.i.d.\ from a distribution $D$ on a sample space $\mathcal Z$. Let $\ell:\R^d\times\mathcal Z\to\R$ be a loss function, and define the population risk
$
\mathsf L(\theta)=\E_{\zeta\sim D}\ell(\theta,\zeta).
$
We consider inference for $\theta^\star=\theta^\star(D)$, defined as
$
\theta^\star=\argmin_{\theta\in\R^d}\mathsf L(\theta),
$
where the minimizer is assumed to be unique.

We consider a streaming data setting, where the algorithm has access to an i.i.d.\ stream rather than to a fixed finite dataset. In our notation, one may view this as $n=\infty$, so that the full-data target equals the population parameter:
$
\theta_n=\theta_\infty=\theta^\star.
$
The finite-data regime can be represented similarly by taking $D$ to be the empirical distribution over a fixed dataset $\mD_n$, in which case the corresponding algorithmic target is the empirical solution $\theta_n=\theta_n(\mD_n)$.

In the streaming setting, we approximate $\theta^\star$ iteratively using a stochastic optimization algorithm.
Let $\htheta_t$ be the parameter estimate at time $t$. 
Starting from a deterministic initial state, say $\htheta_0=0$, 
one popular algorithm to update $\htheta_t$ is stochastic gradient descent \citep{robbins1951stochastic},
$
\htheta_{t+1}
=
\htheta_t-\eta_t \nabla_\theta\ell(\htheta_t,\zeta_t),
$
when the loss is differentiable in $\theta$.

However, our analysis is not limited to stochastic gradient descent. 
More generally it includes any algorithms whose outputs have a limiting distribution, i.e., fluctuate in a predictable way around the true solution, even if the specific distribution is not known to the user of the algorithm. 
As an example,
consider a stochastic heavy ball method, 
which updates a momentum variable $v_t$ along with the parameter estimate $\htheta_t$ and can be written as
\[
v_{t+1}=(1-c\gamma_t)v_t+\gamma_t h(\htheta_t,\zeta_t),
\qquad
\htheta_{t+1}=\htheta_t-\gamma_t v_{t+1},
\]
for some learning rate $(\gamma_t)_{t\ge 1}$ and fixed damping multiplier $c>0$, where $h$ is a stochastic gradient or estimating function; in the differentiable empirical-risk case, $h(\theta,\zeta)=\nabla_\theta\ell(\theta,\zeta)$.

We then consider the $m$th iterate as the output of the randomized algorithm,
$
\htheta_m=\mA_m(\mD_n,S_{m,n}).
$
Here, $\mA_m$ denotes the algorithm run for $m$ iterations, while $S_{m,n}$ encodes the random stream $(\zeta_t)_{0\le t\le m-1}$ together with any additional algorithmic randomization. Only the final parameter iterate $\htheta_m$ needs to be stored for inference.

We also run $K_{m,n}+1$ additional independent copies of the same stochastic optimization algorithm. Each copy starts from the same deterministic initial state, for example $(\widehat\theta_0,v_0)=(0,0)$, and uses its own independent sample path $(\zeta_t^{(k)})_{t=0}^{b-1}$, taking only $b$ steps. On appropriate computational architectures, these copies can be run in parallel, so this step need not require a great deal of additional waiting time. Their outputs $\htheta_{b,i}$ for $i=0,\ldots,K_{m,n}$ are stored for inference.
\end{example}

\section{Inference via sub-randomization}\label{sec:subrand}

In this section, we develop a statistical inference method for the population parameter $\theta^\star$. 
As explained before, 
we also study inference for the finite-sample target $\theta_n$ as an intermediate step.  
Our sub-randomization procedure first approximates the distribution of the randomized estimator around $\theta_n$, and then uses this approximation as a building block for inference on $\theta^\star$. 
Recall that
we observe 
the algorithmic output 
$Z_{m,n}$ and compute an estimator $\hat\theta_m$.
We also generate 
$K_{m,n}+1$ i.i.d.~auxiliary datasets
$Z_{b,n,i}$, $i=0,1, \ldots, K_{m,n}$ and compute auxiliary estimators $\htheta_{b,i}$.

We first define error distributions conditional on the realized dataset $\mD_n$, for which the only randomness comes from the randomized algorithms.
For this we assume that we have 
studentizing scaling matrices $\widehat T_{m,n}=\widehat T_{m,n}(Z_{m,n})\in\Spp$ and $\widehat T_{b,n,0}=\widehat T_{b,n,0}(Z_{b,n,0})\in\Spp$---constructed in a case-by-case manner---such that the distribution of $\widehat T_{m,n}(\htheta_m-\theta_n)$ and the conditional distribution of $\widehat T_{b,n,0}(\htheta_{b,1}-\theta_n)$ given $\mD_n$ and $Z_{b,n,0}$ 
can be controlled, as explained below.
 As a special case, these scaling matrices may be deterministic and may be 
data-independent. 
For instance, when the estimators have a square-root
rate of convergence, we may take
\(
\hT_{m,n}=m^{1/2}I_d\)
and
\(\hT_{b,n,0}=b^{1/2}I_d.
\)
When $\hT_{m,n}=\hat t_m I_d$ is a scalar multiple of the identity, we
abbreviate it as $\hT_{m,n}:=\hat t_m$; and similarly for $\hT_{b,n,0}$.

\subsection{Inference on the full-data target}

Our first result shows that we can perform asymptotic inference when the two conditional laws
\[
\widehat J_{m,n}=\Law\big(\widehat T_{m,n}(\htheta_m-\theta_n)\mid \mD_n\big)
\qquad\text{and}\qquad
\widetilde J_{b,n}
=
\Law\big(
\widehat T_{b,n,0}(\widehat\theta_{b,1}-\theta_n)
\,\big|\,
\mD_n,Z_{b,n,0}
\big)
\]
both converge in probability to a common limiting error distribution $J$. 
The distribution $J$
does not need to be known to the statistician,
and hence other standard approaches, such as pivotal inference, do not directly apply in general.

The crucial definition is the following 
\emph{sub-randomization distribution}, 
which
approximates the conditional law of 
the error
$\widehat T_{b,n,0}(\widehat\theta_{b,1}-\theta_n)$ 
made at the $b$-scale
given
$\mD_n$ and $Z_{b,n,0}$ by the empirical distribution of the empirical 
``pseudo-errors"
\(
\widehat T_{b,n,0}\big(\widehat\theta_{b,i}-\widehat\theta_m\big), i=1,\ldots,K_{m,n}.
\)
Notice that the unknown $\theta_n$ 
from the true error was replaced by the observed $\widehat\theta_m$; this replacement is justified if it induces an error of a smaller order than that of the fluctuations of $\widehat\theta_{b,1}-\theta_n$.

\begin{definition}[The sub-randomization distribution]
Write
\(
\widehat T_{m,n}:=\widehat T_{m,n}(Z_{m,n})\)
and
\(\widehat T_{b,n,0}:=\widehat T_{b,n}(Z_{b,n,0}).
\)
Define the sub-randomization distribution $L_{b,m,n}$ by
\begin{equation}\label{eq:L}
L_{b,m,n}(\Xi)
=
\frac{1}{K_{m,n}}
\sum_{i=1}^{K_{m,n}}
\1\left\{
\widehat T_{b,n,0}\big(\widehat\theta_{b,i}-\widehat\theta_m\big)\in\Xi
\right\},
\qquad
\ \text{for}\ 
\Xi\subseteq\R^d\ \text{measurable}.
\end{equation}
\end{definition}

 Our confidence sets will be 
 scaled and shifted versions of a fixed set, where the scaling is determined to ensure coverage.
Fix a nonempty closed convex set $\mathcal F\subseteq\R^d$ with
$0\in\mathcal F$. 
For a $d\times d$ matrix $M$,
write
$M\cdot\mathcal F=\{Mf:f\in\mathcal F\}$
for the set $\mathcal F$ scaled by $M$.
For a scalar $x$, write analogously
$x\cdot\mathcal F=\{xf:f\in\mathcal F\}$
for the set $\mathcal F$ scaled by $x$.

For a probability distribution
$J$ on $\R^d$, define the quantile $c(1-\alpha)
$ and the scaled set $\Xi_J$ as
\[
c(1-\alpha)
=
\inf\{x\ge0:J(x\cdot\mathcal F)\ge1-\alpha\},
\qquad
\ \text{and}\quad
\Xi_J=c(1-\alpha)\cdot\mathcal F.
\]
Similarly,
for the sub-randomization distribution $L_{b,m,n}$ from
\eqref{eq:L},
define the associated quantile
\begin{equation}\label{eq:cbmn}
c_{b,m,n}(1-\alpha)
=
\inf\big\{x\ge0:L_{b,m,n}(x\cdot\mathcal F)\ge1-\alpha\big\},
\end{equation}
and set
\(
\Xi_{L,b,m,n}=c_{b,m,n}(1-\alpha)\cdot\mathcal F.
\)
Finally, 
define the confidence set $C_{m,n}$
by shifting and scaling $\Xi_{L,b,m,n}$ (and thus $\mathcal{F}$) as
\(
C_{m,n}:=\widehat\theta_m-\widehat T_{m,n}^{-1}\Xi_{L,b,m,n},
\)
whenever $\widehat T_{m,n}$ is invertible.\footnote{When $\widehat T_{m,n}$ is not invertible, we use the pseudoinverse $\widehat T_{m,n}^{\dagger}$ instead of $\widehat T_{m,n}^{-1}$.}
See Algorithm~\ref{alg:subrandomization}.

We say that the $(1-\alpha)$ $\mathcal F$-quantile of $J$ is \emph{regular} if,
writing $c=c(1-\alpha)$, we have $c<\infty$,
$J(\Xi_J)=1-\alpha$, $J(\partial\Xi_J)=0$---where $\partial\Xi_J$ denotes the boundary of the set $\Xi_J$---and, for every $\eta>0$,
\(
J(\max(c-\eta,0)\cdot\mathcal F)
<
1-\alpha
<
J((c+\eta)\cdot\mathcal F).
\)
The following result, proved in Section S.1 of the Supplementary Material (SM), shows that $C_{m,n}$ is an asymptotically valid
$(1-\alpha)$ confidence set for $\theta_n$ under the stated conditions.

\begin{algorithm}[t]
\caption{\footnotesize Sub-randomization confidence set}
\label{alg:subrandomization}
\begin{algorithmic}[1]
\footnotesize
\Require Dataset $\mathcal D_n$ accessed through randomized algorithms; scales $m,b$ ($b<m$); number of auxiliary runs $K_{m,n}$; confidence level $1-\alpha$; closed convex set $\mathcal F\subseteq\mathbb R^d$ with $0\in\mathcal F$ (e.g., unit ball); randomized algorithms $\mathcal A_m$ and $\mathcal A_b$; estimators $\widehat\theta_m$ and $\widehat\theta_b$; studentizers $\widehat T_{m,n}$ and $\widehat T_{b,n}$.
\Ensure Confidence set $C_{m,n}$ for the parameter of interest.
\State Compute the $m$-scale algorithmic output
$
Z_{m,n}=\mathcal A_m(\mathcal D_n,S_{m,n}),
$
where the algorithmic randomness $S_{m,n}$ is independent of $\mathcal D_n$.
\State Compute the $m$-scale estimator and studentizer
$
\widehat\theta_m=\widehat\theta_m(Z_{m,n}),
\qquad
\widehat T_{m,n}=\widehat T_{m,n}(Z_{m,n}).
$

\For{$i=0,1,\ldots,K_{m,n}$}
    \State Draw $S_{b,n,i}$ independently of $(\mathcal D_n,S_{m,n})$ and independently across $i$.
    \State Compute the $b$-scale algorithmic output
    $
    Z_{b,n,i}=\mathcal A_b(\mathcal D_n,S_{b,n,i}).
    $
    \State Compute the auxiliary estimator
    $
    \widehat\theta_{b,i}=\widehat\theta_b(Z_{b,n,i}).
    $
\EndFor

\State Compute the $b$-scale studentizer from the zeroth auxiliary run:
$
\widehat T_{b,n,0}=\widehat T_{b,n}(Z_{b,n,0}).
$

\For{$i=1,\ldots,K_{m,n}$}
    \State Form the studentized pseudo-error
    $
    W_i=\widehat T_{b,n,0}(\widehat\theta_{b,i}-\widehat\theta_m).
    $
\EndFor

\State Define the sub-randomization distribution
$
L_{b,m,n}(\Xi)
=
\frac{1}{K_{m,n}}
\sum_{i=1}^{K_{m,n}}
\mathbf 1\{W_i\in\Xi\},
\quad
\Xi\subseteq\mathbb R^d.
$

\State Compute the empirical calibration radius
$
c_{b,m,n}(1-\alpha)
=
\inf\{x\ge 0:L_{b,m,n}(x\mathcal F)\ge 1-\alpha\}.
$

\State Set
$
\Xi_{L,b,m,n}=c_{b,m,n}(1-\alpha)\mathcal F.
$

\State Return the confidence set
$
C_{m,n}
=
\widehat\theta_m-\widehat T_{m,n}^{-1}\Xi_{L,b,m,n}.
$
\end{algorithmic}
\end{algorithm}

\begin{theorem}[Sub-randomization inference for $\theta_n$]\label{thm:subrand}
Assume $m,n,b,K_{m,n}\to\infty$. Suppose that, for an absolutely continuous
probability distribution $J$ on $\R^d$,
\begin{equation}\label{cond:subrand}
\widehat J_{m,n}\Rightarrow_P J,
\qquad
\widetilde J_{b,n}\Rightarrow_P J.
\end{equation}
Assume also that the smaller-scale centering error is negligible, in the
sense that, for every $\varepsilon>0$,
\(
\mathbb P\left(
\left\|
\widehat T_{b,n,0}(\widehat\theta_m-\theta_n)
\right\|_2>\varepsilon
\,\middle|\,
\mD_n
\right)
\to_P 0.
\)
Assume that $\widehat T_{m,n}$ is invertible with probability tending to one,
and that the $(1-\alpha)$ $\mathcal F$-quantile of $J$ is regular. Then
$C_{m,n}$ is an asymptotically valid
$(1-\alpha)$ confidence set for $\theta_n$:
\[
\mathbb P\Big(
\theta_n\in C_{m,n}
\,\Big|\,
\mD_n
\Big)
\to_P1-\alpha.
\]
\end{theorem}

\begin{remark}[Choice of the auxiliary scale]
The small-scale condition
\(
\widehat T_{b,n,0}(\widehat\theta_m-\theta_n)=o_P(1)
\)
is essential for the centered empirical distribution in \eqref{eq:L} to estimate the same limiting law as the observed statistic. In the common root-rate case, where $\widehat T_{m,n}\asymp \sqrt m\,I_d$ and $\widehat T_{b,n,0}\asymp \sqrt b\,I_d$, this condition is implied by $b/m\to0$ and 
the 
tightness of the sequence of random variables $\sqrt m(\widehat\theta_m-\theta_n)$. Thus $b$ should grow to infinity but remain asymptotically smaller than $m$.
\end{remark}

 One way to establish the required convergence of the error distributions $\widehat J_{m,n}$ and $\widetilde J_{b,n}$ 
 is to introduce 
 scaling matrices $T_{m,n}=T_{m,n}(\mD_n)\in\Spp$ and $T_{b,n}=T_{b,n}(\mD_n)\in\Spp$.
 Then one may establish
 the corresponding conditional limit laws for $T_{m,n}(\htheta_m-\theta_n)$ and $T_{b,n}(\htheta_{b,0}-\theta_n)$, 
 and prove $\widehat T_{m,n}T_{m,n}^{-1}\to_P I_d$ and $\widehat T_{b,n,0}T_{b,n}^{-1}\to_P I_d$.
The invertibility of $\widehat T_{m,n}$ also follows.

\subsubsection{Inference for one-dimensional parameters without scale separation}

In the univariate case, we can construct a confidence interval without imposing the scale separation condition
\(
\widehat T_{b,n,0}(\widehat\theta_m-\theta_n)=o_P(1);
\)
inspired by ideas from subsampling methods, 
see 
e.g., Remark 2.2.4 of \cite{politis1999subsampling}. 

Let $\widehat t_{m,n}=\widehat t_{m,n}(Z_{m,n})$ and $\widehat t_{b,n,0}=\widehat t_{b,n}(Z_{b,n,0})$ be positive scalar studentizers when $d = 1$. For $x\in\mathbb R$, define
a one-dimensional version of \eqref{eq:L} by
$L_{b,m,n}(x)
:=
\frac1{K_{m,n}}
\sum_{i=1}^{K_{m,n}}
I\left\{
\widehat t_{b,n,0}(\widehat\theta_{b,i}-\widehat\theta_m)\le x
\right\}.
$
For $p\in(0,1)$, define the $p$-th quantile 
$c_{b,m,n}^{\mathrm{ts}}(p)
:=
\inf\left\{x\in\mathbb R:
L_{b,m,n}(x)\ge p
\right\}$.
Then, define the following interval, which uses a correction by $\widehat t_{m,n}-\widehat t_{b,n,0}$, instead of the 
$\widehat t_{m,n}$ that would be implied by the general multi-dimensional case:
\begin{equation}\label{eq:two-sided-corrected-interval}
C_{m,n}^{\mathrm{ts}}
:=
\left[
\widehat\theta_m
-
\frac{c_{b,m,n}^{\mathrm{ts}}(1-\alpha/2)}
{\widehat t_{m,n}-\widehat t_{b,n,0}},
\,
\widehat\theta_m
-
\frac{c_{b,m,n}^{\mathrm{ts}}(\alpha/2)}
{\widehat t_{m,n}-\widehat t_{b,n,0}}
\right].
\end{equation}
Then,
assuming that $\widehat t_{m,n}-\widehat t_{b,n,0}$ is positive with probability tending to one (which holds in our examples of interest) and that the relevant scalar quantiles are regular,
we show in 
Section S.4 of the SM
that 
$C_{m,n}^{\mathrm{ts}}$ 
is an asymptotic two-sided $(1-\alpha)$ confidence interval for $\theta_n$, without requiring scale separation. 
For coverage of $\theta^\star$, the corresponding scalar transfer condition is $(\widehat t_{m,n}-\widehat t_{b,n,0})(\theta_n-\theta^\star)\to_P0$.
The advantage of \eqref{eq:two-sided-corrected-interval} is that the correction removes the need to tune $b$ so that $\widehat t_{b,n,0}(\widehat\theta_m-\theta_n)\to_P 0$.
We will typically use this method in our experiments.

\subsection{Inference on the population parameter}
\label{sec:infer_population_parameter}

 We now move to our main goal, 
inference on the population parameter $\theta^\star$.
The scaled error $\widehat T_{m,n}(\htheta_m-\theta^\star)$ can be decomposed as the algorithmic error $\widehat T_{m,n}(\htheta_m-\theta_n) $ and the sampling error $\widehat T_{m,n} (\theta_n-\theta^\star)$.
 If the sampling error $\widehat T_{m,n}(\theta_n-\theta^\star)$ is negligible 
compared to the algorithmic error---which is of unit order---then a confidence set constructed for $\theta_n$ also covers $\theta^\star$ asymptotically. The next theorem formalizes this transfer argument.

\begin{theorem}[Coverage for $\theta^\star$]\label{thm:transfer}
Assume the conditions of \Cref{thm:subrand}. Suppose further that
\(
\widehat T_{m,n}(\theta_n-\theta^\star)\to_P0.
\)
Then
$C_{m,n}$ is an asymptotically valid
$(1-\alpha)$ confidence set for $\theta^\star$:
\(
\mathbb P(\theta^\star\in C_{m,n})\to1-\alpha.
\)
\end{theorem}

This claim, which represents our main correctness result, 
implies that sub-randomization achieves valid asymptotic coverage for the population parameter $\theta^\star$. See Section S.2 of the SM for its proof.
It remains to show that the required conditions apply to 
problems of interest. 
In this, a first observation is the remark below. 

\begin{remark}[Why $m/n\to0$ often implies coverage]
Suppose $\theta_n-\theta^\star=O_P(n^{-1/2})$, as is typical for a
full-data estimator. Suppose also that
\(
\hT_{m,n}=m^{1/2}I_d,
\)
and that the finite-target inference conditions have been verified, for
example with $b/m\to0$ in the usual root-rate case. 
Then
\(
\widehat T_{m,n}(\theta_n-\theta^\star)
=
\sqrt m(\theta_n-\theta^\star)
=
O_P(\sqrt{m/n}).
\)
Hence $m/n\to0$ implies
$\widehat T_{m,n}(\theta_n-\theta^\star)\to_P0$, so the confidence set constructed
for $\theta_n$ also covers $\theta^\star$ asymptotically.
\end{remark}

\begin{remark}
In the univariate case,
the adjusted interval 
from \eqref{eq:two-sided-corrected-interval}
also remains valid, 
without the scale separation condition assumed in \Cref{thm:subrand}, see Section S.4 of the SM.
\end{remark}

\subsection{Connecting sub-randomization with subsampling}

\begin{figure}
\centering
\begin{tikzpicture}[
  x=1.35cm,y=1.25cm,
  >=Latex,
  font=\small,
  est/.style={inner sep=1pt},
  alg/.style={text=green!50!black},
  note/.style={text=blue!70!black, align=left},
  arr/.style={->, thick},
  darr/.style={->, thick, dashed}
]

\node[anchor=west] at (-0.8,3.85) {\textbf{sub-randomization}};

\node[est] (b1) at (0,2.7) {$\htheta_{b,i}$};
\node[est] (m1) at (3,2.7) {$\htheta_{m}$};
\node[est] (n1) at (6,2.7) {$\theta_{n}$};
\node[est] (s1) at (9,2.7) {$\theta^\star$};

\draw[arr] (s1) -- node[alg, above] {$P_{\theta^\star}$} (n1);
\draw[arr] (n1) -- node[alg, above] {$\mathcal A_m$} (m1);
\draw[arr] (n1) to[bend right=28] node[alg, below] {$\mathcal A_{b}$} (b1);

\node[anchor=west] at (-0.8,1.55) {\textbf{subsampling}};

\node[est] (b2) at (0,1) {$\htheta_{b,i}$};
\node[est] (m2) at (3,1) {$\htheta_{m}$};
\node[est] (n2) at (6,1) {$\theta_{n}$};
\node[est] (s2) at (9,1) {$\theta^\star$};

\draw[arr] (s2) -- node[alg, above] {$P_{\theta^\star}$} (n2);
\draw[arr] (n2) -- node[alg, above] {$\mathcal A_m$} (m2);
\draw[arr] (m2) -- node[alg, above] {$\mathcal A_b$} (b2);

\coordinate (midSn) at ($(s2)!0.5!(n2)$);
\node[alg] at ($(midSn)+(0,-0.55)$) {sampling};

\coordinate (midNm) at ($(n2)!0.5!(m2)$);
\node[alg] at ($(midNm)+(0,-0.55)$) {copy};

\coordinate (midMb) at ($(m2)!0.5!(b2)$);
\node[alg] at ($(midMb)+(0,-0.55)$) {sampling};

\end{tikzpicture}
\caption{\footnotesize Relationship between sub-randomization and subsampling. } 
\label{fig:sub-rand-vs-subsampling}
\end{figure}

To help put our work in context, 
we now aim to connect sub-randomization with subsampling, which is one of the most general and powerful methodologies of asymptotic statistical inference. 
It is helpful to study the following triple of 
scaled algorithmic errors 
(at scales $b$ and $m$), along with the scaled sampling error.
Below, $V_n$ is a scaling factor for the sampling error, included to allow comparing the various errors on the same scale: 
\begin{equation}\label{sub-rand-triple-distri}
\left(\hT_{b,n}(\htheta_b - \theta_n),
\hT_{m,n}(\htheta_m - \theta_n),
V_{n}(\theta_n - \theta^\star) \right).
\end{equation}
The first two components describe algorithmic fluctuations,
with marginal limiting law $J$. This coincides with the conditional law assumed in \eqref{cond:subrand}, provided that $J$ does not depend on $\mD_n$. 
 For the third component, 
 it is often the case that the estimators $(\theta_n)_{n\ge 1}$
 have a limiting distribution 
 $J_{\mathrm{sam}}$, 
 i.e., 
 \(
V_n(\theta_n-\theta^\star)\Rightarrow J_{\mathrm{sam}}.
\)
For instance, in the least squares
problem from Example~\ref{ex1},
under suitable
regularity conditions,
the error $\theta_n-\theta^\star$ has an
asymptotic normal distribution $J_{\mathrm{sam}}$ with a specific data-dependent variance after appropriate scaling.
However,
$J_{\mathrm{sam}}$
may differ
from the algorithmic limiting law $J$, 
because it is determined by an entirely different mechanism, namely the sampling process instead of the randomized algorithm.

This setup allows us to clearly compare sub-randomization with subsampling; see Figure~\ref{fig:sub-rand-vs-subsampling} for an illustrative diagram. In classical
subsampling, $n$ is the sample size of the observed data.
There is no separate scale $m$, i.e., 
$\mA_m$ may be viewed as the copy map that returns 
$\mA_m(\mD_n)=\mD_n$, while we can also take
$\htheta_m=\theta_n$. 
Moreover, $\mA_b$ 
takes random subsamples of size $b$ from the full dataset $\mD_n$, and $\htheta_b$ is the analog of the estimator $\theta_n$ computed on those subsamples.
Since $\htheta_m=\theta_n$, the second component of the triple from
\eqref{sub-rand-triple-distri} is identically zero.
The relevant subsampling comparison is therefore between the fluctuation of the statistic computed on a smaller subsample and the sampling fluctuation of the full-sample statistic.
Thus, we consider
the tuple
\begin{equation}\label{subsamp_fullsamp_joint}
\left(
\hT_{b,n}(\htheta_b-\theta^\star),
V_n(\theta_n-\theta^\star)
\right),
\end{equation}
where the two components have marginal limiting distributions $J$ and $J_{\mathrm{sam}}$.\footnote{The first component in \eqref{subsamp_fullsamp_joint} can be centered in either $\theta^\star$ or $\theta_n$ when the induced centering error is negligible. Indeed,
$\widehat T_{b,n}(\htheta_b-\theta^\star)
=
\widehat T_{b,n}(\htheta_b-\theta_n)
+
\widehat T_{b,n}V_n^{-1}V_n(\theta_n-\theta^\star)$.
Hence the two centerings are asymptotically equivalent if
$\|\widehat T_{b,n}V_n^{-1}\|\to0$. This is achieved in many settings, e.g., when root-$n$ asymptotics hold, so that $\widehat T_{b,n}$ grows as $b^{1/2}$, $V_n$ grows as $n^{1/2}$, and $b/n\to0$.}

The success of classical subsampling requires that the subsample statistic
and the full-sample statistic have the same limiting law, i.e.,
$J=J_{\mathrm{sam}}$, and the scaling factors $V_n$ and $\hT_{b,n}$ are
equivalent after proper normalization.
However, the two distributions need not be equal
 for the more general case considered in the present paper, where $\mathcal{A}_m$ is not just a copy map. 
First, dependence or heterogeneity in the data may
prevent a subsample from reproducing the full-sample sampling law. Second, if
$\htheta_b$ is not produced by simple sampling but by another randomized
procedure, such as random projection or stochastic optimization, then its
fluctuations 
are not necessarily related to
the sampling fluctuations of $\theta_n$.
Consequently, as we will see in the experiments, classical subsampling may
fail to achieve the desired coverage when applied naively in the general setting of two-stage observation models.

In contrast,
sub-randomization 
bases inference directly on the algorithmic law. 
Instead of requiring the auxiliary statistic $\widehat\theta_b$ to mimic the sampling fluctuation of $\theta_n$, it only requires the algorithmic fluctuations at scales $b$ and $m$ to share the same limiting law after their respective normalizations.
This yields inference for the full-data target $\theta_n$ without relying on the classical subsampling requirement $J=J_{\mathrm{sam}}$.

\subsection{Sub-randomization inference under converging scale}
\label{cs}

We now consider a special case of the above general setup that we call the
setting of \emph{converging scale}. Suppose that, for some known positive
sequence $(\tau_k)_{k\ge1}$, the
studentizing 
scaling matrices $\widehat T_{k,n}$ ($k\in\{m,b\})$
are asymptotically equivalent 
to the scalar multiples $\tau_k$ of some fixed matrix.
In this case, the sub-randomization procedure can be implemented  
in a simpler way by using 
the known scalar sequence $\tau_k$.

The construction is analogous to
the general case of sub-randomization. 
For measurable sets $\Xi\subseteq\R^d$, define
the following analogue
$L_{b,m,n}'$
of the sub-randomization distribution from \eqref{eq:L} for this setting: 
\begin{equation}\label{lbm}
L_{b,m,n}'(\Xi)
=
\frac1{K_{m,n}}
\sum_{i=1}^{K_{m,n}}
\1\left\{
\tau_b(\htheta_{b,i}-\htheta_m)\in\Xi
\right\},
\qquad
\ \text{for}\ 
\Xi\subseteq\R^d\ \text{measurable}.
\end{equation}
 Continuing analogously, 
for a closed convex set $\mathcal F\subseteq\R^d$ containing the origin, define
\[
c_{b,m,n}'(1-\alpha)
=
\inf\{x\ge0:L_{b,m,n}'(x\cdot\mathcal F)\ge1-\alpha\},
\qquad
\Xi_{L,b,m,n}'=c_{b,m,n}'(1-\alpha)\cdot\mathcal F.
\]
 Consider 
$d\times d$ scaling matrices
 $T_{m,n}$ and $T_{b,n}$
 and let
\[
J_{m,n}
=
\Law\left(T_{m,n}(\htheta_m-\theta_n)\mid\mD_n\right),
\qquad
J_{b,n}
=
\Law\left(T_{b,n}(\htheta_{b,0}-\theta_n)\mid\mD_n\right).
\]
 The scaling matrices 
are not assumed to be known to the analyst. 
They are introduced in order 
to ensure that the formulation of this case is both expressive and analogous to the more general setting of sub-randomization.

\begin{corollary}[Sub-randomization inference under converging scale]\label{ThconfintS}
Consider a sequence of problems as above. Suppose that
$m,n,b,K_{m,n}\to\infty$ and that $\tau_b/\tau_m\to0$. Assume that, for some
absolutely continuous probability distribution $J$ on $\R^d$,
\(
J_{m,n}\Rightarrow_P J\),
\(J_{b,n}\Rightarrow_P J.
\)
Assume further that
\(
M_{m,n}:=T_{m,n}/\tau_m\to_P M,
\quad
M_{b,n}:=T_{b,n}/\tau_b\to_P M,
\)
for some deterministic matrix $M\in\Spp$.

Define the probability law $J_M$ by
\(
J_M(\Xi)=J(M\Xi),\, \Xi\subseteq\R^d\ \textnormal{measurable},
\)
where $M\Xi=\{Mx:x\in\Xi\}$. Let
\(
c_M(1-\alpha)
=
\inf\{x\ge0:J_M(x\cdot\mathcal F)\ge1-\alpha\}\),
\(\Xi_{J_M}=c_M(1-\alpha)\cdot\mathcal F.
\)
Assume that the $(1-\alpha)$ $\mathcal F$-quantile of $J_M$ is regular. If
\(
\tau_m(\theta_n-\theta^\star)\to_P0,
\)
then
\[
\mathbb P\left(
\theta^\star\in
\htheta_m-\tau_m^{-1}\Xi_{L,b,m,n}'
\right)
\to
1-\alpha.
\]
\end{corollary}

The advantage compared to the general sub-randomization construction is that
the statistician does not need to estimate the 
scaling matrices $T_{k,n}$, $k\in\{b,m\}$. 
The
possibly unknown matrix $M$ appears only in the proof of validity, 
and is not required for the implementation.

In the case of a scalar parameter of interest,
 the condition $\tau_b/\tau_m \to 0$ can be relaxed: it is enough that $\tau_m-\tau_b>0$ for all sufficiently large $m$ and that the scalar transfer condition $(\tau_m-\tau_b)(\theta_n-\theta^\star)\to_P0$ holds.
Similarly to \eqref{eq:two-sided-corrected-interval},
a confidence interval under this weaker condition is given by 
\begin{equation}
\label{politis_two_sided_converging_scale}
C_{m,n}'
:=
\left[
\widehat\theta_m
-
\frac{c_{b,m,n}'(1-\alpha/2)}
{ \tau_{m}-\tau_{b}},
\,
\widehat\theta_m
-
\frac{c_{b,m,n}'(\alpha/2)}
{\tau_{m}-\tau_{b}}
\right],
\end{equation}
where for $p \in (0,1)$, $c_{b,m,n}'(p)
=
\inf\{x\in \R:L_{b,m,n}'\big((-\infty, x]\big)\ge p\}$.

\section{Examples}
\label{sec:examples}

We now proceed to illustrate our methodology through two examples: 
least squares regression with correlated noise 
and
stochastic optimization with momentum.

\subsection{Least squares regression with correlated noise}
\label{sec:ls_unif}
In this section, we focus on least squares regression
with a general noise covariance. 
We illustrate how sub-randomization is robust to
noise covariance, whereas subsampling may yield inaccurate coverage rates. 
The trade-off lies in precision: 
sub-randomization yields confidence intervals at the root-$m$ rather than the typical root-$n$ scale.
However, 
sub-randomization provides asymptotic coverage guarantees even when the noise exhibits complex correlation structures. 
Moreover, sub-randomization 
is substantially faster,
enabling the analysis of much larger datasets.

\subsubsection{Inference on the mean with correlated noise}\label{sec:mean_corre_noise}
As a warmup and for clarity, we 
start by introducing the example of inference for the mean. 
Let the
full data 
$Y_n=(Y_{n,1},\dots,Y_{n,n})^\top$
consist of $n$ scalar observations
\[
Y_{n,i}=\theta^\star+u_{n,i},
\qquad i=1,\dots,n,
\]
where $\theta^\star \in \mathbb{R}$
is the fixed unobserved mean. Let
$u_n=(u_{n,1},\dots,u_{n,n})^\top \sim \N(0,\Omega_n)$ be mean-zero Gaussian noise with $n\times n$ covariance matrix
$\Omega_n$. 
We assume that the marginal variances are asymptotically
homogeneous, in the sense that
\(
\max_{1\le i\le n}\left|\operatorname{Var}(u_{n,i})-\sigma^2\right|\to0
\)
for some $\sigma^2>0$; however, we do \emph{not} assume the covariance terms are zero.

Define the full-sample mean $\theta_n = \bar Y_n = n^{-1}\sum_{i=1}^n Y_{n,i}$.
Its variance is $1_n^\top \Omega_n 1_n/n^2$. In general, $\Omega_n$ has
$n(n+1)/2$ unknown entries, and, without additional structure, this variance
cannot be consistently estimated from a single realization of 
$Y_n$. Therefore, classical pivotal inference for $\theta^\star$ cannot generally be used.

Now fix a subsample size $b$ with $b \to \infty$ and $b/n \to 0$. Let $I_{b,n}$ be a uniformly chosen subset of $\{1,\dots,n\}$ of size $b$, and define the subsample mean $\hat\theta_b = b^{-1}\sum_{i \in I_{b,n}} Y_{n,i}$. Then
$\hat\theta_b - \theta_n
= \frac{1}{b} \sum_{i \in I_{b,n}} (u_{n,i} - \bar u_n)$
where 
$\bar u_n = n^{-1}\sum_{i=1}^n u_{n,i}.
$
Conditional on the realized dataset $\mD_n$, the subsampling fluctuation is
determined only by the finite-population variance of the centered errors.
It follows from the finite-population central limit theorem for the more general case of linear regression (\Cref{prop:unif-correlated-clt}) that, under mild regularity conditions,\footnote{
Specifically, if the marginal variances converge uniformly to
$\sigma^2$, $\|\Omega_n\|=O(1)$, and $\log n \ll b$.
}
\(
\Law\bigl(\sqrt b(\hat\theta_b-\theta_n)\mid\mD_n\bigr)
\Rightarrow_P
\mathcal N(0,\sigma^2).
\)

It turns out that subsampling is valid only in very limited 
settings. 
It also follows from \Cref{prop:unif-correlated-clt} that
we have the asymptotic normality result
 \[
\Bigl(1_n^\top \Omega_n 1_n/n^2\Bigr)^{-1/2}
(\theta_n-\theta^\star)
\Rightarrow
\mathcal N(0,1).
\]
Thus the two variances (after rescaling) are $\sigma^2$ and
$1_n^\top \Omega_n 1_n/n = n^{-1}\sum_{i=1}^n\Omega_{n,ii}+n^{-1}\sum_{i\neq j}\Omega_{n,ij}$.
Under the marginal homogeneity assumption, the latter is asymptotically
$\sigma^2+n^{-1}\sum_{i\neq j}\Omega_{n,ij}$.
 Ordinary subsampling is only valid if the average cross-covariance term asymptotically cancels.
For instance,  
if $\Omega_n = \sigma^2 I_n$
and 
$b/n \to 0$, the two distributions in \eqref{subsamp_fullsamp_joint} coincide, where we can choose $\hT_{b,n}= \sqrt{b}$ and $V_n = \sqrt{n}$.

Under correlated noise, however, ordinary subsampling estimates the
empirical variance of the centered realized observations, which
converges to $\sigma^2$ under the above conditions.
In contrast, the sampling distribution of $\theta_n$ is governed by $1_n^\top\Omega_n1_n/n$,
which can differ from $\sigma^2$. 
Consequently, under correlated noise, standard subsampling 
is not guaranteed to achieve coverage.

Sub-randomization avoids this mismatch because it does not try to estimate the sampling law of $\theta_n-\theta^\star$. Instead, it estimates the conditional algorithmic law of a randomized estimator around the full-data target $\theta_n$. 
To summarize the procedure:
let $I_{m,n}$ be an independent uniform subset of size $m$, and set
$
\hat\theta_m=m^{-1}\sum_{i\in I_{m,n}}Y_i.
$
Generate independent auxiliary subsets $I_{b,n,1},\dots,I_{b,n,K_{m,n}}$ of size $b$, and let
$
\hat\theta_{b,i}=b^{-1}\sum_{j\in I_{b,n,i}}Y_j.
$
When $m/n\to0$, $b/m\to0$, $b/\log n\to\infty$, and $K_{m,n}\to\infty$, it follows from \Cref{ThconfintS} that sub-randomization provides an asymptotically valid
confidence interval for the population mean.

Next,  
we 
provide 
a more concrete example where
correlated noise arises.
\begin{example}
[A mean model with a vanishing common shock]\label{example:mean_common_shock}
Consider
$
Y_i=\theta^\star+g_n+\varepsilon_i,\quad i=1,\dots,n,
$
where $\varepsilon_1,\dots,\varepsilon_n$ are independent $\mathcal{N}(0,\sigma^2)$ variables, $g_n\sim \mathcal{N}(0,\tau^2/n)$ is a common shock independent of the $\varepsilon_i$s, and $\sigma^2>0$, $\tau^2>0$ are fixed constants. 
The common shock $g_n$ is small at the individual observation level, since $g_n=O_P(n^{-1/2})$. Nevertheless, it is not negligible for root-$n$ inference about $\theta^\star$.

For the full-data sample mean $\theta_n = \bar{Y}_n$, 
$
\sqrt n(\theta_n-\theta^\star)
=
n^{-1/2}\sum_{i=1}^n\varepsilon_i+\sqrt n g_n
\Rightarrow
\mathcal{N}(0,\sigma^2+\tau^2).
$
On the other hand, ordinary subsampling only estimates $\sigma^2$, as $g_n$ cannot be distinguished from the deterministic mean $\theta^\star$.
This explains why ordinary subsampling does not achieve correct coverage for $\theta^\star$; see Section S.7 of the SM for the full details. 
Since the scale of $g_n$ is much smaller than the width of the 
sub-randomization confidence intervals, sub-randomization can still succeed.
\end{example}

\subsubsection{Least squares with correlated noise}
\label{lsc}

 We now move to our main least squares example. 
We consider inference on the full vector $\theta^\star$ in the regression model from \eqref{lm_model}.
Throughout this section, $d \le n$ is fixed, $\mathbf{X}_n\in\R^{n\times d}$ is deterministic with full column rank, and
\[
y_n=\mathbf{X}_n\theta^\star+u_n.
\]
Let
\[
\theta_n=(\mathbf{X}_n^\top \mathbf{X}_n)^{-1}\mathbf{X}_n^\top y_n,\qquad
H_n=\mathbf{X}_n(\mathbf{X}_n^\top \mathbf{X}_n)^{-1}\mathbf{X}_n^\top,\qquad
Q_n=I_n-H_n,
\]
and write
\(
\epsilon_n=Q_n y_n=Q_n u_n
\)
for the full-sample residual vector. For a vector $v\in\R^n$, let
$\operatorname{Diag}(v)$ denote the diagonal matrix with diagonal $v$. For a
square matrix $A$, let $\operatorname{diag}(A)$ denote the vector of its
diagonal entries. For an index set $I\subset\{1,\dots,n\}$ of size $k \in \{1, \ldots, n\}$, let $\mathbf{X}_{n,I}$ and $y_{n,I}$ denote the selected rows and, when $\mathbf{X}_{n,I}^\top \mathbf{X}_{n,I}$ is invertible, define the sampled least squares estimator
\(
\hat\theta_{k,I}=(\mathbf{X}_{n,I}^\top \mathbf{X}_{n,I})^{-1}\mathbf{X}_{n,I}^\top y_{n,I}.
\)
When $I$ is sampled uniformly without replacement, we simply write $\hat\theta_k$; under the assumptions below, the sampled Gram matrix is invertible with probability tending to one.

Define the ``diagonal sandwich" covariance matrix $\Sigma_n$, which also turns out to be the algorithmic covariance matrix induced by sub-randomization: 
\begin{equation}\label{ls_conditional_cov}
\Sigma_n
=
n(\mathbf{X}_n^\top \mathbf{X}_n)^{-1}
\mathbf{X}_n^\top \operatorname{Diag}(\epsilon_n\odot\epsilon_n) \mathbf{X}_n
(\mathbf{X}_n^\top \mathbf{X}_n)^{-1}.
\end{equation}
The following proposition characterizes the algorithmic covariance of the subsample estimator and also the sampling covariance of the full-sample estimator.
 This is a crucial step towards establishing the validity of sub-randomization. 
See Section S.8 of the SM for the proof.

\begin{proposition}[Uniform sampling under correlated Gaussian noise]\label{prop:unif-correlated-clt}
Assume the following conditions:
\begin{enumerate}[label=(\roman*), ref=(\roman*)]
\item \label{cond:design}
The sequence of designs satisfies
$n^{-1}\mathbf X_n^\top\mathbf X_n\to\Psi$ for some $\Psi\in\Spp$ and
$\max_{1\le i\le n}\|x_{n,i}\|=O(1)$.

\item \label{cond:unif_ls_gaussian_noise}
$u_n\sim\mathcal N(0,\Omega_n)$, where $\|\Omega_n\|=O(1)$.

\item \label{cond:unif_ls_b}
For each $k\in\{m,b\}$, $k/\log n\to\infty$ and $k/n\to0$.
\end{enumerate}
Assume also that the ``mixed sandwich" covariance matrix
\[
G_n
:=
n(\mathbf X_n^\top\mathbf X_n)^{-1}
\mathbf X_n^\top \operatorname{Diag}(\operatorname{diag}(Q_n\Omega_nQ_n))\mathbf X_n
(\mathbf X_n^\top\mathbf X_n)^{-1}
\]
satisfies, for some constants $0<c<C<\infty$,
\(
cI_d\preceq G_n\preceq CI_d
\)
for all sufficiently large $n$.

Then the following hold:
\begin{enumerate}[label=(\alph*), ref=(\alph*)]
\item \label{prop:full_ols}
If
\(
(\mathbf X_n^\top\mathbf X_n)^{-1}
\mathbf X_n^\top\Omega_n\mathbf X_n
(\mathbf X_n^\top\mathbf X_n)^{-1}
\)
is nonsingular for all sufficiently large $n$, then the OLS
estimator satisfies
\(
V_n(\theta_n-\theta^\star)\Rightarrow\mathcal N(0,I_d),
\)
where
\(
V_n=
\left[
(\mathbf X_n^\top\mathbf X_n)^{-1}
\mathbf X_n^\top\Omega_n\mathbf X_n
(\mathbf X_n^\top\mathbf X_n)^{-1}
\right]^{-1/2}.
\)

\item \label{prop_corr_b}
For each $k\in\{m,b\}$,
and with the  algorithmic covariance matrix $\Sigma_n$ from \eqref{ls_conditional_cov},
\[
\Law\left(
k^{1/2}(1-k/n)^{-1/2}\Sigma_n^{-1/2}
(\widehat\theta_k-\theta_n)
\,\middle|\,
\mD_n
\right)
\Rightarrow_P
\mathcal N(0,I_d).
\]

\item \label{prop_cov_consist}
The diagonal and mixed  sandwich covariance matrices satisfy
\(
\Sigma_n-G_n\to_P0.
\)
\end{enumerate}
\end{proposition}

Part~\ref{prop_corr_b} of \Cref{prop:unif-correlated-clt} gives a direct
pivotal route to inference if $\Sigma_n$ can be computed. However,
computing $\Sigma_n$ requires full-data operations involving $\mathbf X_n$. 
Our focus is instead on
settings where full-data calculations are avoided and fast randomized
computation is essential.

As a consequence of parts~\ref{prop_corr_b} and~\ref{prop_cov_consist} of \Cref{prop:unif-correlated-clt}, the randomized estimator satisfies the conditional limit law
\[
\Law\left(
k^{1/2}(1-k/n)^{-1/2}G_n^{-1/2}(\hat\theta_k-\theta_n)
\ \middle|\ \mD_n
\right)\Rightarrow_P \mathcal N(0,I_d).
\]
Classical subsampling attempts to approximate the sampling distribution of the full-sample estimator $\sqrt{n}(\theta_n - \theta^\star)$, whose variability is characterized by the 
``full sandwich" covariance matrix
\(
nV_n^{-2} = n(\mathbf{X}_n^\top \mathbf{X}_n)^{-1}\mathbf{X}_n^\top \Omega_n \mathbf{X}_n (\mathbf{X}_n^\top \mathbf{X}_n)^{-1}.
\)
The validity of this approximation relies on the alignment of the algorithmic (mixed sandwich) and sampling (full sandwich) covariance matrices, given by $G_n$ and $nV_n^{-2}$, and specifically that $G_n-nV_n^{-2} \to 0_{d\times d}$.

We next give a more concrete regression example, which we will also use in our simulations, 
to illustrate the issue.
\begin{example}
Let observations have group labels $\ell_i\in\{1,\dots,K\}$, and write
\[
Y_i=Z_i^\top\theta_0+\alpha_{\ell_i}+u_{n,i},\qquad i=1,\dots,n,
\]
where $Z_i\in\mathbb R^q$ are covariates and $\alpha_{\ell_i}$ is a fixed group effect. Let $\mathbf{Z}_n$ be the matrix with rows $Z_i^\top$, 
and let $M_n\in\mathbb R^{n\times K}$ be the group membership matrix with $i$th row being the indicator $e_{\ell_i}^\top$. 
With $\mathbf{X}_n=(\mathbf{Z}_n,M_n)$ and $\theta^\star=(\theta_0^\top,\alpha^\top)^\top$, the model is thus
$Y=\mathbf{X}_n\theta^\star+u_n$.

The noise $u_n$ has zero  mean,
and may contain both idiosyncratic noise and group-level noise. We write it 
as
$
u_n=\Omega_n^{1/2}\eta_n$,
where
$\E \eta_n=0$,
$\E(\eta_n\eta_n^\top)=I_n,
$
where $\Omega_n$ is the  $n\times n$ covariance matrix of the regression noise. In the Gaussian case covered by \Cref{prop:unif-correlated-clt}, one may take $\eta_n\sim\mathcal N(0,I_n)$.
We consider the covariance model 
$
\Omega_n=I_n+M_nB_nM_n^\top,
$
where $B_n\in\mathbb R^{K\times K}$ is positive semidefinite and is scaled so that $\|\Omega_n\|=O(1)$. 
This is an extension of the common shock model in Example~\ref{example:mean_common_shock}.
The identity term $I_n$
represents idiosyncratic observation-level noise. The low-rank term $M_nB_nM_n^\top$ represents random disturbances shared by observations in correlated groups. 

It turns out that in this setting, even in the two-group case $K=2$, we do not necessarily have
$G_n-nV_n^{-2} \to 0_{d\times d}$.
Thus, subsampling can fail to achieve coverage.

\end{example}

The next proposition shows that sub-randomization remains valid. See Section S.10 of the SM for the proof.
The tradeoff is that sub-randomization leads to a wider interval 
than one based on the full-sample estimator.  
The interval length can be of order $O(1/\sqrt{m})$,  
which is the natural root-$m$ rate for an estimator computed at computational scale $m$.
The advantages lie in the computational gains and general applicability regardless of the covariance structure.

\begin{proposition}[Sub-randomization remains valid under correlated noise]\label{prop:unif-correlated-subrand}
Assume the conditions of \Cref{prop:unif-correlated-clt}.
Assume in addition that $m/n\to0$, $K_{m,n}\to\infty$, and $G_n\to G$ for some $G\in\Spp$. 
For $b<m$,
let
\(
I_{m,n},\ I_{b,n,0},\ I_{b,n,1},\dots,I_{b,n,K_{m,n}}
\)
be mutually independent uniform subsets of $\{1,\dots,n\}$ of sizes $m$ and $b$, independent of the data. Let
\[
\widehat\theta_m=\widehat\theta_{m,I_{m,n}},
\qquad
\widehat\theta_{b,i}=\widehat\theta_{b,I_{b,n,i}},
\qquad
i=0,1,\dots,K_{m,n}.
\]
\begin{enumerate}
\item 
If also $b/m\to0$,  
set $\tau_k=k^{1/2}$ for $k\in\{b,m\}$ in \eqref{lbm} and
let $\mathcal F\subseteq\R^d$ be the Euclidean unit ball.
Then the confidence set from 
\Cref{ThconfintS} is asymptotically valid for $\theta^\star$.
\item If the inferential target is the fixed linear contrast $c^\top \theta^\star$ for a given nonzero vector $c \in \mathbb{R}^d$, define
\(
L'_{b,m,n}(x)
=
\frac{1}{K_{m,n}}
\sum_{i=1}^{K_{m,n}}
\mathbf{1}
\bigl\{
\sqrt{b}\,\bigl(c^\top \widehat{\theta}_{b,i} - c^\top \widehat{\theta}_m\bigr) \le x
\bigr\},
\)
and, for $p\in(0,1)$, let
\(
c'_{b,m,n}(p)=\inf\{x\in\R:L'_{b,m,n}(x)\ge p\}.
\)
Then the interval
\[
\left[
 c^\top\widehat\theta_m-
\frac{c'_{b,m,n}(1-\alpha/2)}{\sqrt m-\sqrt b},
\;
 c^\top\widehat\theta_m-
\frac{c'_{b,m,n}(\alpha/2)}{\sqrt m-\sqrt b}
\right]
\]
attains the nominal asymptotic coverage.
\end{enumerate}
\end{proposition}

\subsubsection{Simulated data examples}
We illustrate sub-randomization using least squares regression with correlated noise. 
We first generate an $n \times 5$ design matrix whose first four columns contain independent standard normal entries. The last column is constructed as a binary vector with ones in the first half of the rows and zeros in the remainder. The matrix is then standardized to obtain the final design matrix.

We generate the noise vector $u_n \in \mathbb{R}^n$ from a multivariate Gaussian distribution $\mathcal{N}(0, \Omega_n)$, 
with covariance 
matrix $\Omega_n = V B V^\top + \eta I_n$.  The matrix $V\in\mathbb{R}^{n\times 2}$ is defined by normalized block indicators.
First, $V_{i1}=1/\sqrt{n_1}$ for $i\le n_1$ and $V_{i1}=0$ for $i>n_1$;
second,
$V_{i2}=0$ for $i\le n_1$ and $V_{i2}=1/\sqrt{n_2}$ for $i>n_1$. We set $n = 10{,}000$ with equal block sizes $n_1 = n_2 = 5,000$. Set $B=\begin{pmatrix}4&2\\2&6\end{pmatrix}$, and $\eta = 1$. 
Computationally, $u_n$ is
generated as $u_n = V g + \varepsilon_n$, where $g \sim \mathcal{N}(0, B)$ represents the shared group-level shocks and $\varepsilon_n \sim \mathcal{N}(0, \eta I_n)$ denotes the independent observation-specific noise.

We consider inference on  $\theta_5$, corresponding to the fixed group effect. 
The confidence interval is constructed using 
sub-randomization 
\eqref{politis_two_sided_converging_scale}.
We set $K_{m,n}=100$
and conduct $500$ Monte Carlo trials 
for each choice of block size $b$ in subsampling and each pair $(b, m)$ in sub-randomization.

Table~\ref{tab:subrandomization-sbm} summarizes the empirical coverage and
average interval lengths across a range of $(b,m)$. 
The coverage of our proposed sub-randomization method remains close to the nominal level in nearly all configurations.
The shortest intervals among the displayed configurations occur
for larger $m$ and smaller $b$, illustrating the trade-off between
computational scale and inferential precision.

For comparison, Table~\ref{tab:subsampling-sbm-uniform-vs-mb} reports results for two
 well-established baseline methods:
classical uniform subsampling and moving block subsampling with block size $b$;
see Section S.5 of the SM for the complete details.
At smaller values of $b$, both  methods exhibit substantial undercoverage. As $b$ increases, uniform subsampling tends toward over-coverage, whereas moving-block subsampling continues to exhibit undercoverage.
 Thus in this particular example, our proposed method outperforms existing methods. 

\begin{table}[ht]
\centering
\caption{\footnotesize Empirical coverage and confidence interval length for $\theta_5$ using sub-randomization, over 
500 Monte Carlo trials for each $(m,b)$ pair.
Each cell reports the coverage probability on the first line (nominal level 0.90) and the average interval length on the second line. Standard deviations of the interval lengths are shown in parentheses.}
\label{tab:subrandomization-sbm}
\footnotesize
\begin{tabular}{rccccc}
\toprule
& \multicolumn{5}{c}{Auxiliary Scale $b$} \\
\cmidrule(lr){2-6}
Scale $m$ & 20 & 40 & 60 & 80 & 100 \\
\midrule
200  & 0.906 & 0.900 & 0.906 & 0.888 & 0.906 \\
     & 0.373 (0.032) & 0.433 (0.038) & 0.515 (0.047) & 0.621 (0.055) & 0.782 (0.067) \\
400  & 0.912 & 0.892 & 0.888 & 0.902 & 0.892 \\
     & 0.231 (0.022) & 0.245 (0.023) & 0.268 (0.024) & 0.295 (0.025) & 0.326 (0.029) \\
600  & 0.896 & 0.890 & 0.908 & 0.886 & 0.912 \\
     & 0.181 (0.017) & 0.186 (0.016) & 0.196 (0.018) & 0.210 (0.018) & 0.223 (0.019) \\
800  & 0.906 & 0.880 & 0.894 & 0.892 & 0.878 \\
     & 0.152 (0.013) & 0.154 (0.014) & 0.161 (0.015) & 0.168 (0.015) & 0.178 (0.016) \\
1000 & 0.880 & 0.864 & 0.870 & 0.886 & 0.896 \\
     & 0.133 (0.012) & 0.134 (0.012) & 0.139 (0.011) & 0.144 (0.013) & 0.149 (0.013) \\
\bottomrule
\end{tabular}
\end{table}

\begin{table}[ht]
\centering
\caption{\footnotesize Empirical coverage and confidence interval length for $\theta_5$ using uniform and moving block subsampling, following the same protocol as in Table~\ref{tab:subrandomization-sbm}. }
\label{tab:subsampling-sbm-uniform-vs-mb}
\footnotesize
\setlength{\tabcolsep}{3pt}
\renewcommand{\arraystretch}{1.3}
\begin{tabular}{llcccccccc}
\hline
Method & Metric$\backslash b$ & 50 & 100 & 200 & 400 & 800 & 1600 & 3200 & 6400 \\
\hline 

& Coverage        & 0.634 & 0.654 & 0.620 & 0.646 & 0.728 & 0.782 & 0.868 & 0.982 \\
Uniform & Mean CI length  & 0.0358 & 0.0359 & 0.0373 & 0.0391 & 0.0429 & 0.0487 & 0.0608 & 0.0958 \\
& SD(CI length)   & 0.0030 & 0.0032 & 0.0033 & 0.0034 & 0.0039 & 0.0043 & 0.0053 & 0.0084 \\
\hline

& Coverage        & 0.614 & 0.644 & 0.694 & 0.714 & 0.754 & 0.790 & 0.866 & 0.814 \\
Moving block & Mean CI length  & 0.0364 & 0.0370 & 0.0398 & 0.0439 & 0.0511 & 0.0620 & 0.0834 & 0.1049 \\
& SD(CI length)   & 0.0035 & 0.0043 & 0.0057 & 0.0086 & 0.0134 & 0.0219 & 0.0431 & 0.0613 \\
\hline
\end{tabular}
\end{table}

\subsubsection{An empirical data analysis example}
\label{sec:ipums}
We consider a dataset sourced
from IPUMS \citep{ruggles2022ipums},
which collects demographic information
on several million workers. 
After preprocessing (as detailed in Section S.5 of the SM), 
we construct 
a dataset from the 1940 census with 
$n=17,\!312,\!687$ datapoints
and $d=38$ variables.\footnote{To be clear, a full data analysis of this data set is still manageable on today's hardware. In fact, in order to evaluate the coverage of our methods, we need to be able to compute the full data least squares estimator, so this is necessary. 
Our point here is not that this dataset is too large to analyze directly, but rather that this dataset is a nontrivial example  
to get a sense of the level of computational savings and statistical efficiency achieved. } 
The response variable is log hourly wage, and the covariates are worker characteristics including education, sex, race, age group, and birthplace group, represented through categorical encoding.  
We fix the design matrix and use the
full-sample least squares solution on the original data
as the true population parameter
$\theta^\star$. 
This full-data computation is performed only once. 
The experiments are performed on hardware detailed in Section S.5 of the SM.

 We generate  noise that is cluster-specific across five levels of education: (i) no schooling, (ii) nursery school through ninth grade, 
(iii) grades 10--12, (iv) 1--4 years of college, and 
(v) five or more years of college. We conduct 500 Monte Carlo replications.
In each replication, we generate $Y_i = X_i^\top \theta^\star + u_i$, $i = 1,\dots,n$, where the error term follows an additive group-level structure:
$u_i = \varepsilon_i + \gamma_{\ell_i} g_{\ell_i}$.
Here, $\ell_i \in \{1,\dots,5\}$ denotes the education group of observation $i$, and $g \sim \mathcal N_5(0,B)$ is a random vector representing shared group-level shocks, with $B$ defined below. 
This hierarchical error specification is designed to emulate the intra-cluster correlation common in socioeconomic data, and is also aligned with our correlated least squares example above.

We set $B$ to have diagonal entries of $20$ and off-diagonal entries equal to five, 
and set the scale factor $\gamma_{\ell_i} = 1/\sqrt{n_j}$ for group $j$, where $n_j$ is the group size. The idiosyncratic noise $\varepsilon_i \overset{iid}{\sim} \mathcal{N}(0, \eta)$, for $\eta$ specified below,
is independent of $g$. To ensure the simulation remains empirically grounded, 
we match variances to the empirical data 
by setting $\eta = \widehat{\sigma}^2 - \operatorname{tr}(B)/n$, where $\widehat{\sigma}^2$ is the residual variance from the original full-data OLS fit. This ensures that the total noise variability, $n^{-1}\operatorname{tr}(\operatorname{Cov}(u))$, matches the empirical variance observed in the raw data.
We then run several methods for statistical inference on the population parameters: baseline methods such as uniform subsampling and block subsampling as well as our  proposed sub-randomization method.

{\bf Results.} 
Figure~\ref{fig:edu_coverage} reports the empirical coverage probabilities for the education-level coefficients. ``EDU4'' corresponds to nursery through grade~4; ``EDU13'' through ``EDU16'' refer to the four years of college, respectively; and ``EDU17+'' denotes five or more years of college. 
For our sub-randomization  method, we set $m = 0.05n$ and $b = 0.01m$. 
Uniform and block subsampling are implemented with $b = 0.05n$. The results show that sub-randomization yields coverage probabilities
close to the nominal $0.90$ level across all coefficients. In contrast, the subsampling methods exhibit severe undercoverage, as they fail to capture the intra-group dependence present in the noise.
 The running time of the various methods is discussed below.

To further examine the inferential precision, we consider the contrast $(\theta_{\text{EDU16}}^\star - \theta_{\text{EDU12}}^\star)/4$, which represents the average increase in log hourly wages per additional year of college education, 
with a population value of $0.0978$.
 Table~\ref{tab:contrast_subrandomization} reports the 
 confidence interval obtained 
 using sub-randomization. The interval width is approximately 0.01, which is small relative to the true parameter value of 0.0978, indicating that 
sub-randomization is informative.

{\bf Ablation: Choice of $b$.}
 Each of the methods we consider 
 requires choosing the small subsample size $b$. 
 We now investigate the 
 effects of this choice. 
Table~\ref{tab:contrast_subsampling} reports the results for the subsampling methods for several choices of $b$, with $b=0.001n = 17,\!312$, $b=0.01n = 173,\!126$, and $b=0.05n = 865,\!634$.
Here, even as $b$ increases, both uniform and block subsampling continue to suffer from undercoverage.
 While the full-data benchmarks can produce shorter confidence
intervals---reflecting the expected trade-off between computational scale $m$
and statistical precision $n$---this gain is spurious given their failure to
achieve nominal coverage.\footnote{ 
The nonmonotonicity in the block subsampling
interval widths reflects the interaction between the block construction and
the highly unbalanced education group structure in the IPUMS design.}
Moreover, sub-randomization is an order of magnitude faster than full-data
subsampling.\footnote{On our hardware,
a full-data least-squares fit takes 38.3 seconds on average, with a standard deviation of 1.5 seconds
across 500 replications.}

\begin{table}[htbp]
\centering
\caption{\footnotesize Sub-randomization inference for the contrast variable $(\theta_{\text{EDU16}}^\star-\theta_{\text{EDU12}}^\star)/4$, with  $m = 0.05 n$ and $K_{m,n}=100$. 
The table reports the running time, confidence interval width, and coverage achieved for various choices of $b$. Results are based on 500 Monte Carlo trials. The nominal coverage level is 0.90. Parentheses
contain standard deviations for running time and width, and $95\%$
Clopper-Pearson confidence intervals for coverage.}
\small
\begin{tabular}{lccc}
\hline
Metric & $b/m{=}0.005$ & $b/m{=}0.01$ & $b/m{=}0.05$ \\
\hline
\addlinespace[0.4em]
Time (sec) & 2.496 (0.129) & 3.054 (0.168) & 7.465 (0.361) \\
\addlinespace[0.4em]
Width & 0.01082 (0.00052) & 0.01008 (0.00244) & 0.00914 (0.00082) \\
\addlinespace[0.4em]
Coverage & 0.892 (0.860, 0.919) & 0.908 (0.878, 0.933) & 0.874 (0.841, 0.903) \\
\hline
\end{tabular}
\label{tab:contrast_subrandomization}
\end{table}

\begin{table}[htbp]
\centering
\caption{\footnotesize Subsampling inference for the contrast variable $(\theta_{\text{EDU16}}^\star-\theta_{\text{EDU12}}^\star)/4$. The same Monte Carlo protocol as in Table~\ref{tab:contrast_subrandomization} is used.}
\footnotesize
\setlength{\tabcolsep}{5pt}
\renewcommand{\arraystretch}{1.30}
\resizebox{0.97\linewidth}{!}{
\begin{tabular}{lcccc}
\hline\noalign{\vskip 2pt}
Method & Metric & $b/n{=}0.001$ & $b/n{=}0.01$ & $b/n{=}0.05$ \\
\hline\noalign{\vskip 2pt}
Uniform subsamp.
& Time (s) & 40.416 (1.574) & 69.768 (3.501) & 234.566 (9.385) \\
& Width & 0.00165 (0.00014) & 0.00176 (0.00015) & 0.00202 (0.00018) \\
& Coverage & 0.582 (0.537,0.626) & 0.618 (0.574,0.661) & 0.694 (0.652,0.734)\\
\hline\noalign{\vskip 2pt}
Block subsamp.
& Time (s) & 39.450 (1.536) & 60.910 (3.054) & 180.789 (7.037) \\
& Width & 0.00335 (0.00020) & 0.01171 (0.00017) & 0.00201 (0.00034) \\
& Coverage & 0.506 (0.461,0.551) & 0.772 (0.732,0.809) & 0.676 (0.633,0.717) \\
\hline\noalign{\vskip 2pt}
\end{tabular}
}
\label{tab:contrast_subsampling}
\end{table}

 \subsection{Stochastic optimization with momentum}
\label{subsec:momentum}
In this section, we illustrate sub-randomization
 for stochastic optimization, specifically for
the stochastic heavy ball method.
 Stochastic optimization
 is widely used in large-scale data analysis and machine learning.
 Over the last decade methods have been developed to use the outputs of stochastic optimization algorithms for inference on population parameters \citep[e.g.,][]{toulis2017asymptotic,fang2018bootstrap,chen2020statistical,zhu2023online}.
 However, these methods are typically designed for specific algorithms such as stochastic gradient descent and do not readily transfer to other algorithms which can have some practical advantages, such as momentum-based optimization.
 Our goal here is to illustrate how our methods can be used seamlessly with a specific example of momentum-based stochastic optimization.  For comparison, we also report the behavior of SGD-iterate inference comparators, including random scaling \citep{lee2022fast}, HiGrad \citep{su2023higrad}, online covariance estimation \citep{zhu2023online}, fixed-$b$ batch means \citep{zhu2021constructing}, and cheap online bootstrap \citep{lam2026cheap}, after porting their implementations to the same squared-hinge example.

We first recall the
general stochastic approximation framework \citep{robbins1951stochastic} in which we seek a root of
$\theta\mapsto H(\theta):=\E_{\zeta\sim D} h(\theta,\zeta)$, where $D$ is a data-generating distribution (or sampling mechanism) and $h$ is an estimating function.
Let $\theta^\star$ satisfy $H(\theta^\star)=0$.
A generic stochastic approximation recursion takes the form
$\widehat\theta_{t+1}=\widehat\theta_t-\eta_t h(\widehat\theta_t,\zeta_t)$,
where $(\eta_t)_{t\ge0}$ is a stepsize schedule and $(\zeta_t)_{t\ge0}$ is a data stream, typically i.i.d.\ from $D$ or
distributed as a Markov chain with stationary distribution $D$.

Empirical risk minimization is a canonical instance of this root-finding framework.
If there is a loss $\ell$ such that $h(\theta,\zeta)=\nabla_\theta \ell(\theta,\zeta)$,  then the recursion becomes SGD:
$\widehat\theta_{t+1}=\widehat\theta_t-\eta_t\nabla_\theta \ell(\widehat\theta_t,\zeta_t)$,
which
aims to
find
a stationary point of $\ell(\theta)=\E_{\zeta\sim D} \ell(\theta,\zeta)$.
This includes SGD-type algorithms for obtaining general $M$-estimators,
for which there are already several methods to use the iterates to obtain statistical inference on $\theta^\star$.
Our goal is to go beyond these algorithms to momentum-based optimization.

\subsubsection{Example: stochastic heavy ball method}

Optimization methods with momentum
are widely used because they can substantially accelerate optimization
 compared to standard stochastic gradient descent.
 They include methods such as Polyak's Heavy Ball \citep{polyak1964some}, Nesterov acceleration \citep{nesterov1983method}, and the highly popular Adam method \citep{Kingma2015adam}.
 Recent work has studied  stochastic versions of these methods,
including stochastic heavy ball methods, stochastic Nesterov acceleration, and variants of Adam; see \cite{Barakat_Bianchi_2021,Barakat_Bianchi_Hachem_Schechtman_2021}.

Our goal is to leverage such stochastic optimization algorithms with momentum to form confidence sets for the true population optimizer.
Starting from $v_0=0$ and $\widehat\theta_0=0$,
an update of a representative
stochastic
heavy ball scheme can be written as
\begin{equation}\label{eq:heavy-ball}
v_{t+1}=(1-c\gamma_t)v_t+\gamma_t\nabla_\theta \ell(\widehat\theta_t,\zeta_t),
\qquad
\widehat\theta_{t+1}=\widehat\theta_t-\gamma_t v_{t+1},
\qquad
 t\ge0,
\end{equation}
with i.i.d.\ data $(\zeta_t)_{t\ge0}$ and a fixed damping multiplier $c>0$.
A typical learning rate schedule is
$\gamma_t=\gamma_0/(t+1)^a$, $t\ge0$, for some $a\ge1/2$.
Under suitable regularity conditions, CLTs for stochastic optimization with
momentum
\citep{Barakat_Bianchi_2021,Barakat_Bianchi_Hachem_Schechtman_2021}
imply a limit of the form
\(
\gamma_t^{-1/2}(\widehat\theta_t-\theta^\star)
\Rightarrow
\mathcal N(0,\Sigma),
\)
where $\theta^\star$ is a critical point of
\(\mathsf L(\theta)=\E\ell(\theta,\zeta)\) and $\Sigma$ depends on the
data distribution, the loss, and the algorithmic hyperparameters.
 Thus,
 the asymptotic covariance $\Sigma$
 of the limiting Gaussian law
is not known to the user,
and pivot-based inference
methods remain elusive.
We consider applying sub-randomization for inference based on the
stochastic heavy ball method.

The crucial observation is that sub-randomization only requires knowing that there \emph{is} a limiting distribution and a known normalizing scale, not
\emph{what} the covariance matrix or other nuisance parameters in that limiting distribution are.
Since the limiting distribution applies with the known scale $\tau_k = \gamma_k^{-1/2}$, we are in the setting of converging scale from \Cref{cs}.
We can consider
both streaming and finite-data regimes.
In a streaming regime, $D$ is the population distribution and $\theta^\star$ is the population parameter.
In a finite-data regime, one may take $D$ to be the empirical distribution over a fixed dataset $\mD_n$, in which case the finite-data target is the empirical solution $\theta_n=\theta_n(\mD_n)$ and inference on $\theta^\star$ proceeds via the transfer step in Theorem~\ref{thm:transfer} once the
sampling error is negligible at the algorithmic scale.

Sub-randomization inference can be used as follows.
Run the stochastic
heavy ball method \eqref{eq:heavy-ball} once for $m$ steps to obtain
$\widehat\theta_m$.
 Using independently sampled data (from the true population in the streaming setting, or from the empirical distribution in the finite data setting),
run the same algorithm $K_m$ times for
$b<m$ steps to obtain
$\widehat\theta_{b,1},\ldots,\widehat\theta_{b,K_m}$. Define
 the sub-randomization distribution from \eqref{lbm} by
\[
L_{b,m}^{\mathrm{HB}}(\Xi)
=
\frac1{K_m}\sum_{i=1}^{K_m}
\1\left\{
\gamma_b^{-1/2}(\widehat\theta_{b,i}-\widehat\theta_m)\in\Xi
\right\},
\qquad
\Xi\subseteq\R^d\ \textnormal{measurable}.
\]
For a fixed closed convex set $\mathcal F\subseteq\R^d$ containing zero, let
\(
c_{b,m}^{\mathrm{HB}}(1-\alpha)
=
\inf\{x\ge0:L_{b,m}^{\mathrm{HB}}(x\cdot\mathcal F)\ge1-\alpha\}\),
\(\Xi_{L,b,m}^{\mathrm{HB}}
=
c_{b,m}^{\mathrm{HB}}(1-\alpha)\cdot\mathcal F.
\)
The resulting confidence set is
\begin{equation}\label{eq:hb-ci-stream}
C_m^{\mathrm{HB}}(1-\alpha)
=
\widehat\theta_m
-
\gamma_m^{1/2}\Xi_{L,b,m}^{\mathrm{HB}}.
\end{equation}
Under the regularity conditions ensuring the CLT
from \cite{Barakat_Bianchi_2021,Barakat_Bianchi_Hachem_Schechtman_2021}
and the scale separation condition
\(
\frac{\gamma_b^{-1/2}}{\gamma_m^{-1/2}}
=
\left(\frac{\gamma_m}{\gamma_b}\right)^{1/2}
\to0,
\)
\Cref{ThconfintS} implies
\(
\mathbb P\{\theta^\star\in C_m^{\mathrm{HB}}(1-\alpha)\}\to1-\alpha.
\)
For the polynomial schedule $\gamma_t=\gamma_0/(t+1)^a$, this scale
separation is implied by $b/m\to0$.

To construct a confidence interval
for $u^\top \theta^\star$,
given a nonzero vector $u \in \R^d$, similarly define
\(
\widetilde L^{\mathrm{HB}}_{b,m}(x)
=
\frac{1}{K_m}
\sum_{i=1}^{K_m}
\mathbf{1}
\bigl\{
\gamma_b^{-1/2}\,\bigl(u^\top \widehat{\theta}_{b,i} - u^\top \widehat{\theta}_m\bigr) \le x
\bigr\},
\)
and, for $p \in (0,1)$, let
\(
c^{\mathrm{HB}}_{b,m}(p)
=
\inf\{x\in \R:\widetilde L^{\mathrm{HB}}_{b,m}(x)\ge p\}.
\)
Then the confidence interval \eqref{politis_two_sided_converging_scale} specializes to
\begin{equation}\label{hb-politis-CI}
\left[
u^\top\widehat\theta_m
-
\frac{c^{\mathrm{HB}}_{b,m}(1-\alpha/2)}{\gamma_{m}^{-1/2}-\gamma_{b}^{-1/2}},
\;
u^\top\widehat\theta_m
-
\frac{c^{\mathrm{HB}}_{b,m}(\alpha/2)}{\gamma_{m}^{-1/2}-\gamma_{b}^{-1/2}}
\right]
\end{equation}
and attains nominal coverage asymptotically without requiring the condition $b/m \to 0$.

\paragraph{Illustration: squared-hinge classification with momentum.}
We investigate the performance of sub-randomization when the heavy ball method
is used to minimize a squared-hinge classification risk. Consider
binary labels $Y\in\{-1,1\}$ and the risk minimization problem
\begin{equation}\label{eq:svm-squared-hinge}
\min_{\theta,\beta_0}
\frac12\E\big(\max\{0,1-Y(\theta^\top X+\beta_0)\}\big)^2.
\end{equation}
We write $\vartheta^\star=((\theta^\star)^\top,\beta_0^\star)^\top$ for the augmented minimizer and focus on
inference for the first coordinate $\theta_1^\star$ of its $\theta$-block. In this example, the stochastic algorithms are run on the augmented parameter $\vartheta=(\theta^\top,\beta_0)^\top$; the reported confidence interval is the projection onto the first coordinate of the $\theta$-block.

We generate $(X,Y)$ from a
 five-dimensional
two-component Gaussian mixture $\sum_{j=1}^2 \phi_j\mathcal{N}(\mu_j,\Sigma)$ with $\phi_1=0.2$, $\phi_2=0.8$, $\mu_1=(1,1,1,0,0)^\top$, $\mu_2=(0,0,1,1,1)^\top$, and $\Sigma=0.5I_5$.
The label $Y$ is $+1$ if the sample comes from the component with mean $\mu_1$ and $-1$ otherwise.

For the heavy ball method, set $\gamma_t=0.4/(t+1)^{0.55}$ and use damping multiplier $c=2$ in \eqref{eq:heavy-ball}.
Additional sensitivity analyses over the damping multiplier and the stepsize exponent are reported in Section S.6 of the SM, showing largely similar results.
We also compare with standard SGD, using the augmented-parameter updates
$\widehat\vartheta_{t+1}=\widehat\vartheta_t-\gamma_t\nabla_\vartheta\ell(\widehat\vartheta_t,\zeta_t)$ for all $t$.
This method also admits a CLT under standard conditions \citep[see e.g.,][]{chen2002stochastic,li2022revisiting} but generally with
an asymptotic covariance matrix different from that for the heavy ball method.
For reference, we additionally compare with random scaling \citep{lee2022fast}, HiGrad \citep{su2023higrad}, the online covariance matrix estimator of \cite{zhu2023online}, the fixed-$b$ batch-means method of \cite{zhu2021constructing}, and the cheap online bootstrap of \cite{lam2026cheap}.  
To isolate differences among the inference
procedures, we hold the underlying SGD dynamics fixed across all  comparators and use the common stepsize schedule
$\gamma_t=0.4/(t+1)^{0.55}$. 

For the method of \cite{zhu2023online}, we run averaged SGD after the common burn-in and estimate the covariance of the averaged iterate by the recursive online batch-means construction.  We report both the full overlapping estimator and the non-overlapping estimator, using the polynomial batch-start sequence $a_r=\lfloor r^{2/(1-a)}\rfloor$ with $a=0.55$ and $C=1$.  For the method from \cite{zhu2021constructing}, we use the scalar marginal fixed-$b$ F-statistic interval with $q=30$ increasing-batch-size batches; 
the critical value is calibrated by Monte Carlo for the same stepsize exponent.  
For the cheap online bootstrap, we maintain exponentially weighted bootstrap trajectories online and report $B=3$ and $B=10$.

We use the sub-randomization confidence interval from \eqref{hb-politis-CI}, setting $K_m=50$, $b=500$, and $m=50{,}000$, and run 500 replications.
To warm-start all sub-randomization methods, within each Monte Carlo replication we initialize the $m$-scale run and all $b$-scale auxiliary runs at the same output of a 1000-iteration burn-in run of standard SGD or heavy ball, depending on which type of randomized algorithm is considered. The post-burn-in data streams are then sampled independently across runs.
All reported comparator methods use the same burn-in convention, and the post-burn-in data stream length is $m+K_m b=75{,}000$.  
We compute the population minimizer $\theta^\star$ by evaluating the gradient of the squared-hinge population risk in \eqref{eq:svm-squared-hinge} using truncated-normal moments under the Gaussian-mixture distribution and numerically solving the resulting deterministic first-order conditions. This yields $\theta_1^\star=0.601076$.


\begin{table}[htbp]
\centering
\caption{\footnotesize
Inference for the first coordinate of the squared-hinge risk minimizer.
Sub-randomization is applied to standard SGD and stochastic
heavy ball with $\gamma_t=0.4/(t+1)^{0.55}$ and heavy-ball damping multiplier
$c=2$. The remaining rows are SGD-iterate comparators ported to the same
squared-hinge setting. All results are based on
500 Monte Carlo replications. }
\label{tab:svm-inference-main}

\small
\setlength{\tabcolsep}{5pt}
\begin{tabular}{@{}lccc@{}}
\hline
\multirow{2}{*}{Method}
    & Coverage
    & 95\% CP interval
    & \multirow{2}{*}{Mean length (MC SE)} \\
    & (nominal 0.90)
    & for coverage
    & \\
\hline
Heavy ball sub-randomization
    & \cellcolor{green!30}0.886
    & \cellcolor{green!30}[0.855, 0.913]
    & \cellcolor{green!30}0.0818 (0.0005) \\

SGD sub-randomization
    & \cellcolor{green!30}0.906
    & \cellcolor{green!30}[0.877, 0.930]
    & \cellcolor{red!30}0.1198 (0.0008) \\

Random scaling \citep{lee2022fast}
    & \cellcolor{red!30}0.778
    & \cellcolor{red!30}[0.739, 0.814]
    & 0.0340 (0.0009) \\

HiGrad \citep{su2023higrad}
    & \cellcolor{red!30}0.676
    & \cellcolor{red!30}[0.633, 0.717]
    & 0.0263 (0.0005) \\

online BM (overlap) \citep{zhu2023online}
    & \cellcolor{red!30}0.478
    & \cellcolor{red!30}[0.433, 0.523]
    & 0.0159 (0.0004) \\

online BM (non-overlap) \citep{zhu2023online}
    & \cellcolor{red!30}0.544
    & \cellcolor{red!30}[0.499, 0.588]
    & 0.0186 (0.0005) \\

BM/F-stat. \citep{zhu2021constructing}
    & \cellcolor{red!30}0.526
    & \cellcolor{red!30}[0.481, 0.571]
    & 0.0191 (0.0007) \\

COnB, $B=3$ \citep{lam2026cheap}
    & \cellcolor{green!30}0.904
    & \cellcolor{green!30}[0.875, 0.928]
    & \cellcolor{red!30}0.0911 (0.0085) \\

COnB, $B=10$ \citep{lam2026cheap}
    & \cellcolor{red!30}0.958
    & \cellcolor{red!30}[0.937, 0.974]
    & \cellcolor{red!30}0.1048 (0.0151) \\
\hline
\end{tabular}
\end{table}

{\bf Results.}
Table~\ref{tab:svm-inference-main} reports 95\% Clopper--Pearson intervals for the coverage of nominal 90\% confidence intervals, along with average interval lengths.
The heavy-ball sub-randomization interval has empirical coverage $0.886$, with 95\% Clopper--Pearson interval $[0.855,0.913]$, and mean length $0.0818$.
The matched SGD sub-randomization interval has empirical coverage $0.906$ and mean length $0.1198$.
Thus, in this setting, heavy-ball sub-randomization yields a confidence interval about $32\%$ shorter than matched SGD while retaining coverage compatible with the nominal level.
Random scaling, HiGrad, the Zhu--Chen--Wu methods, and the Zhu--Dong
fixed-$b$ BM/F-statistic interval, implemented using $q=30$
increasing-batch-size batches, all exhibit undercoverage.

By contrast, the cheap online bootstrap attains nominal coverage for $B=3$, with empirical coverage $0.904$ and mean length $0.0911$, and is conservative for $B=10$, with empirical coverage $0.958$.
However, COnB also exhibits high interval-length variability: for $B=3$, the 
mean length is 0.0911, but the
the 95th quantile 0.3210, reflecting a right-skewed distribution with occasional very long intervals.

\section{Summary and discussion}

In this work, we have proposed a framework
for
statistical inference 
when using randomized algorithms to analyze massive data sets.
 We formalized this through 
two-stage observation models 
that include a standard random sampling step for the data and 
randomized estimators.
 We have developed sub-randomization, a method that relies on running the randomized estimators multiple times, and argued both theoretically and empirically that it can achieve asymptotic coverage. 
In future work, it would be of interest to study broader classes of problems and algorithms where 
our methods could be used, including additional classes of stochastic optimization and adaptive data analysis.

\section*{Acknowledgements}

The authors thank John Duchi, Lucas Janson,
Dimitris Politis,
Jianfeng Yao, and Leda Wang for valuable feedback. 
This work was supported in part by
the NSF and ONR. Zhixiang Zhang was partially supported by NSFC-12401331, University of Macau MYRG-GRG2024-00260-FST-UMDF and SRG2023-00053-FST.

\section{Appendix}
\subsection{Proof of Theorem \ref{thm:subrand}}
\label{sec:proof_subrand}

\begin{lemma}[Conditional empirical-law LLN with shared randomness]\label{lem:conditional-empirical-law}
Let $\mathcal G_n$ be sigma-fields, let $K_n\to\infty$, and let $W_{n,1},\dots,W_{n,K_n}$ be random vectors in $\R^d$ such that, conditionally on $\mathcal G_n$, they are i.i.d. with common conditional law $\mu_n$. Define the empirical measure
\[
\widehat\mu_n
=
\frac{1}{K_n}\sum_{i=1}^{K_n}\delta_{W_{n,i}}.
\]
If $\mu_n\Rightarrow_P J$ for a deterministic probability measure $J$ on $\R^d$, then $\widehat\mu_n\Rightarrow_P J$.
\end{lemma}

\begin{proof}
Let $f:\R^d\to\R$ be bounded and continuous. Write
\[
\widehat\mu_n(f)
=
\int f\,d\widehat\mu_n
=
\frac{1}{K_n}\sum_{i=1}^{K_n} f(W_{n,i}),
\qquad
\mu_n(f)
=
\int f\,d\mu_n
=
\mathbb E[f(W_{n,1})\mid \mathcal G_n].
\]
Conditionally on $\mathcal G_n$, the variables $f(W_{n,1}),\dots,f(W_{n,K_n})$ are i.i.d., hence
\(
\mathbb E[\widehat\mu_n(f)\mid \mathcal G_n]=\mu_n(f)
\)
and
\[
\mathrm{Var}(\widehat\mu_n(f)\mid \mathcal G_n)
=
\frac{1}{K_n}\mathrm{Var}(f(W_{n,1})\mid \mathcal G_n)
\le
\frac{\|f\|_\infty^2}{K_n}.
\]
Therefore
\[
\mathbb E\big[(\widehat\mu_n(f)-\mu_n(f))^2\big]
=
\mathbb E\big[\mathrm{Var}(\widehat\mu_n(f)\mid \mathcal G_n)\big]
\le
\frac{\|f\|_\infty^2}{K_n}
\to0.
\]
So $\widehat\mu_n(f)-\mu_n(f)\to0$ in probability. Since $\mu_n\Rightarrow_P J$, we also have $\mu_n(f)\to \int f\,dJ$ in probability. Hence
\[
\widehat\mu_n(f)\to \int f\,dJ
\]
in probability. Taking $f$ over a countable bounded-Lipschitz convergence-determining class and using the bounded-Lipschitz metric, we conclude that $\widehat\mu_n\Rightarrow_P J$.
\end{proof}

\paragraph{Proof of \Cref{thm:subrand}.}
Let
\[
\Delta_{b,m,n}
=
\widehat T_{b,n,0}(\widehat\theta_m-\theta_n).
\]
By assumption, $\Delta_{b,m,n}=o_P(1)$ conditionally on $\mD_n$. Conditionally on
\[
\mathcal G_n=\sigma(\mD_n,Z_{m,n},Z_{b,n,0}),
\]
the variables
\[
W_{n,i}
=
\widehat T_{b,n,0}(\widehat\theta_{b,i}-\widehat\theta_m),
\qquad i=1,\ldots,K_{m,n},
\]
are i.i.d. Their conditional law is the law of
\[
\widehat T_{b,n,0}(\widehat\theta_{b,1}-\theta_n)-\Delta_{b,m,n}
\]
given $\mathcal G_n$. Since $\widehat\theta_{b,1}$ is conditionally
independent of $Z_{m,n}$ given $\mD_n$ and $Z_{b,n,0}$, the conditional law of
$\widehat T_{b,n,0}(\widehat\theta_{b,1}-\theta_n)$ given $\mathcal G_n$ is
$\widetilde J_{b,n}$. Therefore, by $\widetilde J_{b,n}\Rightarrow_P J$ and
$\Delta_{b,m,n}=o_P(1)$, the conditional law of
$W_{n,1}$ given $\mathcal G_n$ converges weakly in probability to $J$; for instance, the bounded-Lipschitz distance between a law and its translation by $\Delta_{b,m,n}$ is at most $\|\Delta_{b,m,n}\|$.
The conditional empirical-law lemma gives
\[
L_{b,m,n}\Rightarrow_P J.
\]

By the regularity of the $(1-\alpha)$ $\mathcal F$-quantile of $J$,
\[
c_{b,m,n}(1-\alpha)\to_P c(1-\alpha).
\]
Indeed, the sets $x\cdot\mathcal F$ are nested in $x$ because
$0\in\mathcal F$ and $\mathcal F$ is convex, and the strict inequalities in
the regularity assumption rule out nonunique limiting quantiles.

Let
\[
X_{m,n}=\widehat T_{m,n}(\widehat\theta_m-\theta_n).
\]
By assumption, the conditional law of $X_{m,n}$ given $\mD_n$ converges weakly
in probability to $J$. Since
$c_{b,m,n}(1-\alpha)\to_P c(1-\alpha)$ and $J(\partial\Xi_J)=0$, the usual
portmanteau argument for random continuity sets yields
\[
\mathbb P\left(
X_{m,n}\in\Xi_{L,b,m,n}
\,\middle|\,
\mD_n
\right)
\to_P
J(\Xi_J)
=
1-\alpha.
\]
Finally,
\[
\theta_n\in C_{m,n}
\quad\Longleftrightarrow\quad
\widehat T_{m,n}(\widehat\theta_m-\theta_n)\in\Xi_{L,b,m,n}.
\]
This proves the claim.
\qed

\subsection{Proof of Theorem \ref{thm:transfer}}
\begin{proof}
Let
\[
X_{m,n}=\widehat T_{m,n}(\widehat\theta_m-\theta_n),
\qquad
R_{m,n}=\widehat T_{m,n}(\theta_n-\theta^\star),
\qquad
c_{m,n}=c_{b,m,n}(1-\alpha).
\]
By the proof of \Cref{thm:subrand}, $c_{m,n}\to_P c(1-\alpha)$. Also,
$X_{m,n}\Rightarrow J$ unconditionally, because its conditional law given
$\mD_n$ converges weakly in probability to $J$, and $R_{m,n}\to_P0$ by
assumption. Therefore
\[
(X_{m,n},R_{m,n},c_{m,n})
\Rightarrow
(X,0,c),
\qquad X\sim J,\quad c=c(1-\alpha).
\]
The event $\{\theta^\star\in C_{m,n}\}$ is equivalent to
\[
X_{m,n}+R_{m,n}\in c_{m,n}\cdot\mathcal F.
\]
The boundary of the limiting event is contained in
$\{X\in\partial(c\cdot\mathcal F)\}$, which has probability zero by the
regularity condition in \Cref{thm:subrand}. Hence, by the portmanteau theorem,
\[
\mathbb P(\theta^\star\in C_{m,n})
\to
\mathbb P(X\in c\cdot\mathcal F)
=
1-\alpha.
\]
\end{proof}

\subsection{Proof of Corollary \ref{ThconfintS}}
\label{pfThconfintS}

\begin{proof}
We apply \Cref{thm:subrand} with limiting law $J_M$ and with the deterministic choices
\[
\hT_{m,n}=\tau_m I_d,
\qquad
\hT_{b,n,0}=\tau_b I_d.
\]
Since $T_{m,n}/\tau_m\to_P M$ and $T_{b,n}/\tau_b\to_P M$, Slutsky's theorem gives
\[
\Law\left(\tau_m(\htheta_m-\theta_n)\mid\mD_n\right)
\Rightarrow_P
J_M
\]
and
\[
\Law\left(\tau_b(\htheta_{b,0}-\theta_n)\mid\mD_n\right)
\Rightarrow_P
J_M.
\]
Moreover,
\[
\tau_b(\htheta_m-\theta_n)
=
\frac{\tau_b}{\tau_m}
M_{m,n}^{-1}T_{m,n}(\htheta_m-\theta_n)
=
o_P(1)
\]
conditionally in probability, because $\tau_b/\tau_m\to0$,
$M_{m,n}^{-1}=O_P(1)$, and $T_{m,n}(\htheta_m-\theta_n)=O_P(1)$
conditionally in probability. Thus the smaller-scale centering condition in
\Cref{thm:subrand} holds.

The empirical distribution $L_{b,m,n}'$ is exactly the sub-randomization
distribution from \Cref{thm:subrand} under these scalar choices of
$\hT_{m,n}$ and $\hT_{b,n,0}$. 
Then,
\Cref{thm:subrand} yields
\[
\mathbb P\left(
\theta_n\in
\htheta_m-\tau_m^{-1}\Xi_{L,b,m,n}'
\,\middle|\,
\mD_n
\right)
\to_P
1-\alpha.
\]
The assertion for $\theta^\star$ follows from \Cref{thm:transfer}, because
$\tau_m(\theta_n-\theta^\star)\to_P0$.
\end{proof}

\subsection{Univariate confidence interval without small-scale condition}
\label{app:politis-two-sided}

The construction in \Cref{thm:subrand} is stated for general multivariate
sets. In that setting, centering the auxiliary empirical distribution at
$\widehat\theta_m$ introduces a random translation of the estimated set, and
the small-scale condition is used to make this translation asymptotically
negligible. In the scalar two-sided case, \eqref{eq:two-sided-corrected-interval} gives
an exact correction analogous to Remark~2.2.4 of \cite{politis1999subsampling}.

Throughout this subsection, assume that $d=1$. Let
$\widehat t_{m,n}=\widehat t_{m,n}(Z_{m,n})$ and
$\widehat t_{b,n,0}=\widehat t_{b,n}(Z_{b,n,0})$ be positive scalar
studentizers. Define
\begin{equation*}
\widehat J_{m,n}
=
\mathcal L\left(
\widehat t_{m,n}(\widehat\theta_m-\theta_n)
\,\middle|\,
\mathcal D_n
\right)
\end{equation*}
and
\begin{equation*}
\widetilde J_{b,n}
=
\mathcal L\left(
\widehat t_{b,n,0}(\widehat\theta_{b,1}-\theta_n)
\,\middle|\,
\mathcal D_n,Z_{b,n,0}
\right).
\end{equation*}
We now state the result for the confidence interval defined via \eqref{eq:two-sided-corrected-interval}.

\begin{theorem}[Scalar two-sided correction]\label{thm:politis-two-sided}
Assume $m,n,b,K_{m,n}\to\infty$. Suppose that
\begin{equation*}
\widehat J_{m,n}\Rightarrow_P J,
\qquad
\widetilde J_{b,n}\Rightarrow_P J,
\end{equation*}
where $J$ is a probability distribution on $\mathbb R$ with distribution
function $F$. Let
\begin{equation*}
q_-
=
q(\alpha/2)
=
\inf\{x\in\mathbb R:F(x)\ge\alpha/2\},
\quad
q_+
=
q(1-\alpha/2)
=
\inf\{x\in\mathbb R:F(x)\ge1-\alpha/2\}.
\end{equation*}
Assume that the two quantiles are regular and continuity points of $F$, in
the sense that
$F(q_-)=F(q_- -)=\alpha/2$, and
$F(q_+)=F(q_+ -)=1-\alpha/2,
$
and, for every $\eta>0$,
\begin{equation*}
F(q_- -\eta)<\alpha/2<F(q_-+\eta),
\qquad
F(q_+ -\eta)<1-\alpha/2<F(q_+ +\eta).
\end{equation*}
Assume also that
\begin{equation*}
\mathbb P\left(
\widehat t_{m,n}>\widehat t_{b,n,0}
\,\middle|\,
\mathcal D_n
\right)
\to_P1.
\end{equation*}
Then
\begin{equation*}
\mathbb P\left(
\theta_n\in C_{m,n}^{\mathrm{ts}}
\,\middle|\,
\mathcal D_n
\right)
\to_P
1-\alpha.
\end{equation*}
If, in addition,
\begin{equation*}
(\widehat t_{m,n}-\widehat t_{b,n,0})(\theta_n-\theta^\star)\to_P0,
\end{equation*}
then
\begin{equation*}
\mathbb P\left(
\theta^\star\in C_{m,n}^{\mathrm{ts}}
\right)
\to
1-\alpha.
\end{equation*}
\end{theorem}

\begin{proof}
Define the empirical distribution function
\begin{equation*}
U_{b,m,n}(x)
:=
\frac1{K_{m,n}}
\sum_{i=1}^{K_{m,n}}
\1\left\{
\widehat t_{b,n,0}(\widehat\theta_{b,i}-\theta_n)\le x
\right\},
\qquad x\in\mathbb R,
\end{equation*}
and let
\begin{equation*}
u_{b,m,n}(p)
:=
\inf\{x\in\mathbb R:U_{b,m,n}(x)\ge p\},
\qquad p\in(0,1).
\end{equation*}
Conditionally on $\mathcal H_n=\sigma(\mathcal D_n,Z_{b,n,0}),$
the variables
$\widehat t_{b,n,0}(\widehat\theta_{b,i}-\theta_n),
\; i=1,\ldots,K_{m,n},
$
are i.i.d. with conditional law $\widetilde J_{b,n}$. Hence,
by Lemma \ref{lem:conditional-empirical-law},
\begin{equation*}
U_{b,m,n}\Rightarrow_P J.
\end{equation*}
By regularity of the two quantiles,
\begin{equation}\label{eq:u-quantile-converges-two-sided}
u_{b,m,n}(\alpha/2)\to_P q_-,
\qquad
u_{b,m,n}(1-\alpha/2)\to_P q_+.
\end{equation}

Now set
\begin{equation*}
\Delta_{b,m,n}
:=
\widehat t_{b,n,0}(\widehat\theta_m-\theta_n).
\end{equation*}
For every $x\in\mathbb R$,
\begin{equation*}
L_{b,m,n}(x)
=
U_{b,m,n}(x+\Delta_{b,m,n}).
\end{equation*}
Therefore, by the definition of the generalized inverse, for each $p\in(0,1)$,
\begin{equation}\label{eq:quantile-translation-two-sided}
c_{b,m,n}^{\mathrm{ts}}(p)
=
u_{b,m,n}(p)-\Delta_{b,m,n}.
\end{equation}

On the event $\widehat t_{m,n}>\widehat t_{b,n,0}$,
\begin{align*}
\theta_n\in C_{m,n}^{\mathrm{ts}}
&\Longleftrightarrow
c_{b,m,n}^{\mathrm{ts}}(\alpha/2)
\le
(\widehat t_{m,n}-\widehat t_{b,n,0})(\widehat\theta_m-\theta_n)
\le
c_{b,m,n}^{\mathrm{ts}}(1-\alpha/2)
\\
&\Longleftrightarrow
u_{b,m,n}(\alpha/2)
\le
\widehat t_{m,n}(\widehat\theta_m-\theta_n)
\le
u_{b,m,n}(1-\alpha/2),
\end{align*}
where the second equivalence uses \eqref{eq:quantile-translation-two-sided}.

The right-hand side in the two-sided event contains the infeasible quantiles
$u_{b,m,n}(\alpha/2)$ and $u_{b,m,n}(1-\alpha/2)$, but both are asymptotically
deterministic by \eqref{eq:u-quantile-converges-two-sided}. Moreover,
conditionally on $\mathcal D_n$, the statistic
$\widehat t_{m,n}(\widehat\theta_m-\theta_n)$ is independent of the auxiliary
runs used to construct these quantiles.
Since
\[
\mathcal L\!\left(
\widehat t_{m,n}(\widehat\theta_m-\theta_n)\,\middle|\,\mathcal D_n
\right)\Rightarrow_P J,
\]
the scalar random-threshold argument at the regular quantiles gives
\[
\mathbb P\!\left(
\widehat t_{m,n}(\widehat\theta_m-\theta_n)
\le
u_{b,m,n}(1-\alpha/2)\,\middle|\,\mathcal D_n
\right)\to_P1-\alpha/2,
\]
\[
\mathbb P\!\left(
\widehat t_{m,n}(\widehat\theta_m-\theta_n)
<
u_{b,m,n}(\alpha/2)\,\middle|\,\mathcal D_n
\right)\to_P\alpha/2.
\]
Therefore
\begin{align*}
\mathbb P\!\left(
u_{b,m,n}(\alpha/2)
\le
\widehat t_{m,n}(\widehat\theta_m-\theta_n)
\le
u_{b,m,n}(1-\alpha/2)
\,\middle|\,\mathcal D_n
\right)
\to_P 1-\alpha.
\end{align*}
The event $\widehat t_{m,n}\le\widehat t_{b,n,0}$ has asymptotically
negligible conditional probability, so the asserted conditional coverage for
$\theta_n$ follows.

For the population parameter, define
$Y_{m,n}
:=
\widehat t_{m,n}(\widehat\theta_m-\theta_n)$
and
$R_{m,n}
:=
(\widehat t_{m,n}-\widehat t_{b,n,0})(\theta_n-\theta^\star).$
On the event $\widehat t_{m,n}>\widehat t_{b,n,0}$,
\begin{align*}
\theta^\star\in C_{m,n}^{\mathrm{ts}}
&\Longleftrightarrow
c_{b,m,n}^{\mathrm{ts}}(\alpha/2)
\le
(\widehat t_{m,n}-\widehat t_{b,n,0})(\widehat\theta_m-\theta^\star)
\le
c_{b,m,n}^{\mathrm{ts}}(1-\alpha/2)
\\
&\Longleftrightarrow
u_{b,m,n}(\alpha/2)
\le
Y_{m,n}+R_{m,n}
\le
u_{b,m,n}(1-\alpha/2).
\end{align*}
By assumption, $R_{m,n}\to_P0$. 
Combining this with
\eqref{eq:u-quantile-converges-two-sided}, the weak convergence of
$\widehat t_{m,n}(\widehat\theta_m-\theta_n)$, and the regularity of the
quantile gives
\begin{equation*}
\mathbb P\left(
\theta^\star\in C_{m,n}^{\mathrm{ts}}
\right)
\to
1-\alpha.
\end{equation*}

\end{proof}

When the scalar studentizers are deterministic, say
$\widehat t_{m,n}=t_m$ and $\widehat t_{b,n,0}=t_b$, the interval in
\eqref{eq:two-sided-corrected-interval} becomes
\begin{equation*}
\left[
\widehat\theta_m
-
\frac{c_{b,m,n}^{\mathrm{ts}}(1-\alpha/2)}
{t_{m}- t_{b}},
\,
\widehat\theta_m
-
\frac{c_{b,m,n}^{\mathrm{ts}}(\alpha/2)}
{t_{m}-t_{b}}
\right].
\end{equation*}
Thus the factor $(t_m-t_b)^{-1}$ appears exactly as in the scalar two-sided
subsampling correction of \cite{politis1999subsampling}. The reason this
does not require $t_b/t_m\to0$ is that the proof does not require
$L_{b,m,n}$ itself to consistently estimate the limiting law
$J$; instead, its quantile differs from the infeasible quantile by the exact
translation in \eqref{eq:quantile-translation-two-sided}.

\subsection{Additional details of numerical experiments}
\label{sec:more-simulation-details}
{\bf Moving block subsampling.}
In the numerical experiments in Sections \ref{sec:ls_unif} and \ref{sec:ipums}, moving block subsampling is implemented as follows. For a given block size $b$, each auxiliary replication draws a starting row index
$s \sim \mathrm{Unif}\{1,\ldots,n-b+1\},
$
and forms the consecutive block
$
I_b=\{s,s+1,\ldots,s+b-1\}.
$
The least-squares estimator is then computed using only the observations indexed by $I_b$. This procedure is repeated independently $K$ times to approximate the corresponding subsampling distribution.

{\bf IPUMS experiment computing environment and preprocessing.}
For the IPUMS experiment, the computations were performed on the Wharton HPC3 high-performance computing cluster, which runs Rocky Linux 8 on AWS EC2 compute nodes with shared WekaFS storage.  The software environment used R version 4.5.2 with GCC 12.2.0.

The data are prepared as follows. We use the 1940 U.S. census sample. The outcome is log hourly wage. Age is grouped into 11 categories: under 20, five-year age groups from 20--24 through 55--59, ages 60--69, and ages 70 and above. Birthplace is grouped into the ten most common categories, with the remaining categories combined into an ``Other'' category. Education, sex, race, age group, and birthplace group are included as categorical covariates using the usual dummy-variable coding.  The resulting sample has $n=17,\!312,\!687$ observations, and the final design matrix has $d=38$ columns.

\subsection{Additional ablations for stochastic heavy ball}
\label{app:momentum-ablation}

This section reports additional ablations for the squared-hinge classification experiment in Section~\ref{subsec:momentum}.  The data-generating distribution, target coordinate, warm-start convention, sub-randomization interval, and Monte Carlo replication count are the same as in the main stochastic-heavy-ball experiment.  We vary the heavy-ball damping multiplier $c$ in
\[
v_{t+1}=(1-c\gamma_t)v_t+\gamma_t\nabla_\theta \mathcal{\ell}(\widehat\theta_t,\zeta_t),
\qquad
\widehat\theta_{t+1}=\widehat\theta_t-\gamma_t v_{t+1},
\]
and the stepsize exponent $a$ in $\gamma_t=0.4/(t+1)^a$.  Specifically, we use
\[
c\in\{1,1.5,2,2.5,3\},
\qquad
a\in\{0.50,0.55,0.60\}.
\]
For each configuration we use $m=50{,}000$, $K_{m,n}=50$, $b=500$, burn-in 1000, and 500 Monte Carlo replications.

\begin{table}
\centering
\caption{Damping and exponent ablations for stochastic heavy-ball
sub-randomization in the squared-hinge classification experiment.
The final column divides the mean heavy-ball interval length by the
matched SGD sub-randomization length at the same stepsize exponent.}
\label{tab:momentum-ablation-grid}
\begin{tabular}{ccccccc}
\hline
$a$ & $c$ & SGD cov. & SGD len. & HB cov. & HB len. & HB/SGD \\
\hline
0.50 & 1.0 & 0.866 & 0.1530 & 0.844 & 0.1641 & 1.072 \\
0.50 & 1.5 & 0.866 & 0.1530 & 0.934 & 0.1227 & 0.802 \\
0.50 & 2.0 & 0.866 & 0.1530 & 0.880 & 0.1013 & 0.662 \\
0.50 & 2.5 & 0.866 & 0.1530 & 0.798 & 0.0891 & 0.582 \\
0.50 & 3.0 & 0.866 & 0.1530 & 0.620 & 0.0793 & 0.518 \\
0.55 & 1.0 & 0.906 & 0.1198 & 0.888 & 0.1347 & 1.125 \\
0.55 & 1.5 & 0.906 & 0.1198 & 0.958 & 0.0981 & 0.819 \\
0.55 & 2.0 & 0.906 & 0.1198 & 0.886 & 0.0818 & 0.683 \\
0.55 & 2.5 & 0.906 & 0.1198 & 0.750 & 0.0714 & 0.596 \\
0.55 & 3.0 & 0.906 & 0.1198 & 0.652 & 0.0640 & 0.534 \\
0.60 & 1.0 & 0.944 & 0.0986 & 0.892 & 0.1131 & 1.147 \\
0.60 & 1.5 & 0.944 & 0.0986 & 0.966 & 0.0838 & 0.849 \\
0.60 & 2.0 & 0.944 & 0.0986 & 0.902 & 0.0683 & 0.692 \\
0.60 & 2.5 & 0.944 & 0.0986 & 0.878 & 0.0597 & 0.605 \\
0.60 & 3.0 & 0.944 & 0.0986 & 0.888 & 0.0527 & 0.534 \\
\hline
\end{tabular}
\end{table}

The grid illustrates the trade-off between interval length and coverage. At $a=0.55$, the damping multipliers $c=1$ and $c=2$ yield nearly identical heavy-ball coverages of $0.888$ and $0.886$, respectively. However, $c=2$ produces substantially shorter intervals, with an HB-to-SGD length ratio of $0.683$, compared with $1.125$ for $c=1$. Increasing the damping multiplier further to $c=2.5$ or $c=3$ produces even shorter intervals but leads to substantial undercoverage. At $a=0.60$, the setting $c=2$ achieves empirical coverage of $0.902$ and a length ratio of $0.692$, whereas $c=3$ achieves empirical coverage of $0.888$ and a length ratio of $0.534$. The main text focuses on $a=0.55$ and $c=2$, for which coverage is compatible with the nominal level and the mean interval length is substantially shorter than that of matched SGD.


\subsection{Coverage in the common-shock mean model}
\label{cs2}

The next proposition demonstrates the undercoverage of classical subsampling for the common-shock mean model. See Section \ref{pf:sub_mean} for the proof.

\begin{proposition}
\label{prop:common-shock-mean}
Consider the common-shock model
\[
Y_i=\theta^\star+g_n+\varepsilon_i,\qquad i=1,\dots,n,
\]
where $\varepsilon_1,\dots,\varepsilon_n$ are independent
$\mathcal N(0,\sigma^2)$ variables, $g_n\sim\mathcal N(0,\tau^2/n)$ is
independent of the $\varepsilon_i$'s, and $\sigma^2,\tau^2>0$ are fixed.
Suppose $m,n,b,K_{m,n}\to\infty$, with $m/n\to0$, $b/m\to0$ and $b/\log n \to \infty$.

Let
\[
L^{\mathrm{sr}}_{b,m,n}(A)
=
\frac1{K_{m,n}}
\sum_{j=1}^{K_{m,n}}
\1\{\sqrt b(\widehat\theta_{b,j}-\widehat\theta_m)\in A\},
\qquad A\subseteq\R\ \textnormal{measurable},
\]
and define
\[
c^{\mathrm{sr}}_{b,m,n}(1-\alpha)
=
\inf\{x\ge0:L^{\mathrm{sr}}_{b,m,n}([-x,x])\ge1-\alpha\}.
\]
For the sub-randomization confidence interval
\[
C^{\mathrm{sr}}_{b,m,n}(1-\alpha)
=
\left[
\widehat\theta_m-
\frac{c^{\mathrm{sr}}_{b,m,n}(1-\alpha)}{\sqrt m},
\,
\widehat\theta_m+
\frac{c^{\mathrm{sr}}_{b,m,n}(1-\alpha)}{\sqrt m}
\right],
\]
we have
\[
\mathbb P\!\left(\theta_n\in C^{\mathrm{sr}}_{b,m,n}(1-\alpha)\mid\mD_n\right)
\to_P
1-\alpha
\]
and
\[
\mathbb P\!\left(\theta^\star\in C^{\mathrm{sr}}_{b,m,n}(1-\alpha)\right)
\to
1-\alpha.
\]

For ordinary subsampling, let $K_n\to\infty$ and define
\[
L^{\mathrm{sub}}_{n,b}(A)
=
\frac1{K_n}
\sum_{j=1}^{K_n}
\1\{\sqrt b(\widehat\theta_{b,j}-\theta_n)\in A\},
\]
and
\[
c^{\mathrm{sub}}_{n,b}(1-\alpha)
=
\inf\{x\ge0:L^{\mathrm{sub}}_{n,b}([-x,x])\ge1-\alpha\}.
\]
Then the usual symmetric subsampling interval
\[
C^{\mathrm{sub}}_{n,b}(1-\alpha)
=
\left[
\theta_n-\frac{c^{\mathrm{sub}}_{n,b}(1-\alpha)}{\sqrt n},
\,
\theta_n+\frac{c^{\mathrm{sub}}_{n,b}(1-\alpha)}{\sqrt n}
\right]
\]
satisfies
\[
\mathbb P\!\left(\theta^\star\in C^{\mathrm{sub}}_{n,b}(1-\alpha)\right)
\to
2\Phi\!\left(
z_{1-\alpha/2}\frac{\sigma}{\sqrt{\sigma^2+\tau^2}}
\right)-1
<1-\alpha.
\]
\end{proposition}

\subsubsection{Proof of Proposition \ref{prop:common-shock-mean}}
\label{pf:sub_mean}

First,
$
\theta_n-\theta^\star=\bar\varepsilon_n+g_n.
$
Therefore
$
\sqrt n(\theta_n-\theta^\star)
=
n^{-1/2}\sum_{i=1}^n\varepsilon_i+\sqrt n g_n
\Rightarrow
\mathcal{N}(0,\sigma^2+\tau^2),
$
because the two terms are independent Gaussian random variables with variances $\sigma^2$ and $\tau^2$.

Next consider the conditional law of a uniform subsample mean. Since the common shock $g_n$ is the same for all observations, it disappears after centering:
$
\hat\theta_k-\theta_n
=
\frac1k\sum_{i\in I_{k,n}}(Y_i-\bar Y_n)
=
\frac1k\sum_{i\in I_{k,n}}(\varepsilon_i-\bar\varepsilon_n).
$
The finite-population variance satisfies
$
\frac1n\sum_{i=1}^n(Y_i-\bar Y_n)^2
=
\frac1n\sum_{i=1}^n(\varepsilon_i-\bar\varepsilon_n)^2
\to_P
\sigma^2.
$
Moreover $\max_i|\varepsilon_i-\bar\varepsilon_n|=O_P(\sqrt{\log n})$, so the usual finite-population Lindeberg condition holds whenever $k/\log n\to\infty$ and $k/n\to0$. Hence, conditionally on $\mD_n$,
$
\Law\left(\sqrt k(\hat\theta_k-\theta_n)\mid\mD_n\right)
\Rightarrow_P
\mathcal{N}(0,\sigma^2)
$
for $k=b$ and for $k=m$.

For ordinary subsampling, the conditional empirical law of
$
\sqrt b(\hat\theta_b^{\mathrm{sub}}-\theta_n)
$
therefore converges in probability to $\mathcal{N}(0,\sigma^2)$, by the conditional law of large numbers for the auxiliary subsampling draws. Consequently
$
c^{\mathrm{sub}}_{n,b}(1-\alpha)\to_P z_{1-\alpha/2}\sigma.
$
Combining this with
$
\sqrt n(\theta_n-\theta^\star)\Rightarrow \mathcal{N}(0,\sigma^2+\tau^2)
$
gives
$
\mathbb P\left(\theta^\star\in C^{\mathrm{sub}}_{n,b}(1-\alpha)\right)
\to
\mathbb P\left(
|\mathcal{N}(0,\sigma^2+\tau^2)|\le z_{1-\alpha/2}\sigma
\right),
$
which equals
$
2\Phi\left(
z_{1-\alpha/2}\sigma/\sqrt{\sigma^2+\tau^2}
\right)-1.
$
This is strictly smaller than $1-\alpha$ whenever $\tau^2>0$.

We now prove the sub-randomization claim. The conditional law above gives
$\Law\left(\sqrt m(\hat\theta_m-\theta_n)\mid\mD_n\right)
\Rightarrow_P
\mathcal{N}(0,\sigma^2)
$
and
$
\Law\left(\sqrt b(\hat\theta_b-\theta_n)\mid\mD_n\right)
\Rightarrow_P
\mathcal{N}(0,\sigma^2).
$
Furthermore,
$
\sqrt b(\hat\theta_m-\theta_n)
=
\sqrt{b/m}\sqrt m(\hat\theta_m-\theta_n)
=
o_P(1),
$
because $b/m\to0$ and $\sqrt m(\hat\theta_m-\theta_n)=O_P(1)$. Therefore
$
\sqrt b(\hat\theta_{b,j}-\hat\theta_m)
=
\sqrt b(\hat\theta_{b,j}-\theta_n)
-
\sqrt b(\hat\theta_m-\theta_n)
$
has conditional limiting law $\mathcal{N}(0,\sigma^2)$. Since $K_{m,n}\to\infty$, the empirical law $L^{\mathrm{sr}}_{b,m,n}$ converges in probability to $\mathcal{N}(0,\sigma^2)$, and hence
$
c^{\mathrm{sr}}_{b,m,n}(1-\alpha)\to_P z_{1-\alpha/2}\sigma.
$
Also, conditionally on $\mD_n$,
$
\sqrt m(\hat\theta_m-\theta_n)
\Rightarrow_P
\mathcal{N}(0,\sigma^2).
$
It follows that
$
\mathbb P\left(\theta_n\in C^{\mathrm{sr}}_{b,m,n}(1-\alpha)\mid\mD_n\right)
\to_P
1-\alpha.
$

Finally,
$
\sqrt m(\theta_n-\theta^\star)
=
\frac{\sqrt m}{\sqrt n}\sqrt n(\theta_n-\theta^\star)
=
o_P(1),
$
because $m/n\to0$ and $\sqrt n(\theta_n-\theta^\star)=O_P(1)$. Therefore the confidence interval that has conditional coverage for $\theta_n$ also has unconditional asymptotic coverage for $\theta^\star$. This proves the result.
\qed

\subsection{Proof of Proposition \ref{prop:unif-correlated-clt}}
\label{sec:pf_unif_corr_clt}

We state the following facts regarding finite population CLT  \citep{hajek1960limiting} that will be used below. 

\begin{lemma}[Uniform-sampling LLN and CLT]\label{lem:unif-sampling-clt}
Let $I_{k_n,n}$ be a uniform subset of $\{1,\dots,n\}$ of size $k_n$, independent of the data, with $k_n\to\infty$ and $k_n/n\to0$.
\begin{enumerate}
\item If $a_{n,1},\dots,a_{n,n}\in\R$ are random variables measurable with respect to $\mD_n$ and $n^{-1}\sum_{i=1}^n a_{n,i}^2=O_P(1)$, then
\[
\frac1{k_n}\sum_{i\in I_{k_n,n}} a_{n,i}-\frac1n\sum_{i=1}^n a_{n,i}\to_P0.
\]

\item If $w_{n,1},\dots,w_{n,n}\in\R^d$ are random vectors measurable with respect to $\mD_n$ such that
\[
\sum_{i=1}^n w_{n,i}=0,\qquad
\frac1n\sum_{i=1}^n w_{n,i}w_{n,i}^\top\to_P I_d,
\]
and
\[
\frac{1}{\min(k_n,n-k_n)}\cdot \frac{\max_{1\le i\le n}\|w_{n,i}\|^2}{(n-1)^{-1}\sum_{i=1}^n \|w_{n,i}\|^2}\to_P 0,
\]
then
\[
\Law\left(
\frac{1}{\sqrt{k_n(1-k_n/n)}}\sum_{i\in I_{k_n,n}} w_{n,i}
\ \middle|\ \mD_n
\right)\Rightarrow_P \mathcal N(0,I_d).
\]
\end{enumerate}
\end{lemma}

\begin{proof}
For the first statement, conditionally on $\mD_n$,
\[
\mathrm{Var}\left(\frac1{k_n}\sum_{i\in I_{k_n,n}} a_{n,i}\,\middle|\,\mD_n\right)
=
\frac{1-k_n/n}{k_n(n-1)}
\sum_{i=1}^n (a_{n,i}-\bar a_n)^2,
\qquad
\bar a_n=\frac1n\sum_{i=1}^n a_{n,i}.
\]
Since $n^{-1}\sum a_{n,i}^2=O_P(1)$, the right-hand side is $O_P(k_n^{-1})$, which proves the claim.

For the second statement, fix $t\in\R^d$. Conditionally on $\mD_n$, the values $z_{n,i}=t^\top w_{n,i}$ form a finite population of size $n$ with mean $\bar z_n=\frac1n\sum_i z_{n,i}=0$, finite-population variance $v_n=(n-1)^{-1}\sum_i z_{n,i}^2\to_P t^\top t$, and $m_n=\max_i z_{n,i}^2\le \|t\|^2\max_i\|w_{n,i}\|^2$. The assumed condition implies $\frac{1}{\min(k_n,n-k_n)}\cdot \frac{m_n}{v_n}\to_P 0$. By the univariate finite-population CLT \citep{hajek1960limiting},
\[
\frac{1}{\sqrt{k_n(1-k_n/n)}}\sum_{i\in I_{k_n,n}} z_{n,i}
\Rightarrow_P \mathcal{N}(0,t^\top t).
\]
The vector statement follows from the Cramér--Wold device.
\end{proof}

\noindent \textbf{Proof of Proposition \ref{prop:unif-correlated-clt}.}
Let $h_{n,i}=x_{n,i}^\top(\mathbf X_n^\top\mathbf X_n)^{-1}x_{n,i}$ denote the $i$th leverage score. From Condition \ref{cond:design}, 
 we have 
 $$\max_i h_{n,i}\le \max_i \|x_{n,i}\|^2/\lambda_{\min}(\mathbf{X}_n^\top \mathbf{X}_n)=O(n^{-1}).$$
Part (a) is immediate under the stated nonsingularity assumption, since
\(
\theta_n-\theta^\star=(\mathbf{X}_n^\top \mathbf{X}_n)^{-1}\mathbf{X}_n^\top u_n
\)
and $u_n\sim \mathcal N(0,\Omega_n)$.

We now prove part \ref{prop_corr_b}. The argument is the same for
$k=b$ and $k=m$, so we give it for a generic sequence
$k=k_n\in\{b,m\}$ satisfying the assumptions of the proposition.
Let
\[
\hat G_{k,n}=\frac1k \mathbf{X}_{n,I_{k,n}}^\top \mathbf{X}_{n,I_{k,n}},
\qquad
\bar G_n=\frac1n \mathbf{X}_n^\top \mathbf{X}_n.
\]
 Lemma~\ref{lem:unif-sampling-clt}(1), applied entrywise to $a_{n,i}=x_{n,ir}x_{n,is}$, yields
\(
\hat G_{k,n}-\bar G_n\to_P0
\)
conditionally on $\mD_n$. Therefore
\(
\hat G_{k,n}^{-1}-\bar G_n^{-1}\to_P0
\)
conditionally on $\mD_n$, because $\bar G_n\to\Psi\in\Spp$.

Now
\[
\widehat\theta_k-\theta_n
=
(\mathbf{X}_{n,I_{k,n}}^\top \mathbf{X}_{n,I_{k,n}})^{-1}\mathbf{X}_{n,I_{k,n}}^\top \epsilon_{n,I_{k,n}}
=
\hat G_{k,n}^{-1}\frac1k\sum_{i\in I_{k,n}} x_{n,i}\epsilon_{n,i}.
\]
Let $\Gamma_n = Q_n \Omega_n Q_n$.
Since $\epsilon_n\sim\mathcal N(0,\Gamma_n)$ with $\sup_n\|\Gamma_n\|_{\op}<\infty$, we have
\begin{equation}\label{cond:cond_noise}
\frac1n\sum_{i=1}^n\epsilon_{n,i}^2=O_P(1),
\qquad
\max_{1\le i\le n}|\epsilon_{n,i}|=O_P(\sqrt{\log n}).
\end{equation}
The conditional covariance of $k^{-1/2}\sum_{i\in I_{k,n}} x_{n,i}\epsilon_{n,i}$ is
\(
\frac{1-k/n}{n-1}\cdot\)
\(
\sum_{i=1}^n \epsilon_{n,i}^2 x_{n,i}x_{n,i}^\top,
\)
which is $O_P(1)$ implied by Conditions \ref{cond:design} and the first bound in \eqref{cond:cond_noise}. Consequently,
\begin{equation}\label{equ:hbeta_betan_dif}
\sqrt k\,(\widehat\theta_k-\theta_n)
=
\bar G_n^{-1}\frac1{\sqrt k}\sum_{i\in I_{k,n}} x_{n,i}\epsilon_{n,i}
+
o_P(1).
\end{equation}

Define
\[
w_{n,i}
=
\Sigma_n^{-1/2}\bar G_n^{-1}x_{n,i}\epsilon_{n,i},
\qquad
i=1,\dots,n.
\]
Since $\mathbf{X}_n^\top \epsilon_n=\mathbf{X}_n^\top Q_n y_n=0$, we have $\sum_i w_{n,i}=0$. Moreover,
\[
\frac1n\sum_{i=1}^n w_{n,i}w_{n,i}^\top
=
\Sigma_n^{-1/2}
\bar G_n^{-1}
\Big(\frac1n\sum_{i=1}^n \epsilon_{n,i}^2x_{n,i}x_{n,i}^\top\Big)
\bar G_n^{-1}
\Sigma_n^{-1/2}
=
I_d.
\]
Further,
\begin{equation}\label{eq:max_wni}
\max_i \|w_{n,i}\|
\le
\|\Sigma_n^{-1/2}\|\,
\|\bar G_n^{-1}\|\,
\max_i \|x_{n,i}\|\,
\max_i |\epsilon_{n,i}|.
\end{equation}

By part \ref{prop_cov_consist}, proved below, and the assumed lower bound $cI_d\preceq G_n$, we have $\lambda_{\min}(\Sigma_n)$ bounded away from zero with probability tending to one. Together with Condition \ref{cond:design}, this gives $\|\Sigma_n^{-1/2}\|=O_P(1)$ and $\|\bar G_n^{-1}\|=O(1)$. Therefore \eqref{eq:max_wni} and \eqref{cond:cond_noise} imply
\[
\max_i\|w_{n,i}\|^2=O_P(\log n).
\]
The assumption $k/\log n\to\infty$ is therefore sufficient for the H\'ajek condition in \Cref{lem:unif-sampling-clt}.

Lemma~\ref{lem:unif-sampling-clt}(2) now yields
\[
\Law\left(
\frac{1}{\sqrt{k(1-k/n)}}\sum_{i\in I_{k,n}} w_{n,i}
\ \middle|\ \mD_n
\right)\Rightarrow_P \mathcal N(0,I_d).
\]
Since
\[
\frac{1}{\sqrt{k(1-k/n)}}\sum_{i\in I_{k,n}} w_{n,i}
=
k^{1/2}(1-k/n)^{-1/2}\Sigma_n^{-1/2}\bar G_n^{-1}\frac1k\sum_{i\in I_{k,n}} x_{n,i}\epsilon_{n,i},
\]
and by \eqref{equ:hbeta_betan_dif}, we obtain
\[
\Law\left(
k^{1/2}(1-k/n)^{-1/2}\Sigma_n^{-1/2}(\widehat\theta_k-\theta_n)
\ \middle|\ \mD_n
\right)\Rightarrow_P \mathcal N(0,I_d).
\]

We then prove part \ref{prop_cov_consist}. Write the SVD as $\mathbf{X}_n=U_n\Lambda_n R_n^\top$. 
Then
$\Sigma_n
=
nR_n\Lambda_n^{-1}
U_n^\top \operatorname{Diag}(\epsilon_n\odot\epsilon_n) U_n
\Lambda_n^{-1}R_n^\top,
$
while
\(
G_n
=
nR_n\Lambda_n^{-1}
U_n^\top \operatorname{Diag}(\operatorname{diag}(\Gamma_n)) U_n
\Lambda_n^{-1}R_n^\top.
\)
It is therefore enough to show that
\[
U_n^\top \operatorname{Diag}(\epsilon_n\odot\epsilon_n) U_n
-
U_n^\top \operatorname{Diag}(\operatorname{diag}(\Gamma_n)) U_n
\to_P0.
\]

Let $a_{jk,n}=U_{n,\cdot j}\odot U_{n,\cdot k}$ and
\(
B_{jk,n}=Q_n\operatorname{Diag}(a_{jk,n})Q_n.
\)
Since $Q_n$ is an orthogonal projection,
\[
\|B_{jk,n}\|_F
\le
\|\operatorname{Diag}(a_{jk,n})\|_F
=
\|U_{n,\cdot j}\odot U_{n,\cdot k}\|_2
\le
\|U_{n,\cdot j}\|_\infty\|U_{n,\cdot k}\|_2.
\]
The leverage bound gives
\(
\|U_{n,\cdot j}\|_\infty
\le
\max_{1\le r\le n}h_{n,r}^{1/2}
=
O(n^{-1/2}),
\)
and $\|U_{n,\cdot k}\|_2=1$. Hence $\|B_{jk,n}\|_F=o(1)$.

For Gaussian $u_n\sim\mathcal N(0,\Omega_n)$,
\[
\operatorname{Var}(u_n^\top B_{jk,n}u_n)
=
2\,\operatorname{tr}\big((\Omega_n^{1/2}B_{jk,n}\Omega_n^{1/2})^2\big)
\le
2\|\Omega_n\|^2\|B_{jk,n}\|_F^2
=
o(1).
\]
This proves
\[
U_n^\top\operatorname{Diag}(\epsilon_n\odot\epsilon_n)U_n
-
U_n^\top\operatorname{Diag}(\operatorname{diag}(\Gamma_n))U_n
\to_P0,
\]
and therefore $\Sigma_n-G_n\to_P0$.
\qed

\subsection{Analysis of the grouped regression example}
\label{sec:a-cle}

Focus on the two-group case $K=2$ and let
\(
x=M_ne_2,
\)
so that $x_i=1\{\ell_i=2\}$ and $n_2=x^\top x$ is the number of observations in group two.
The covariance term entering the full-sample sandwich covariance contains
\(
x^\top\Omega_nx
=
n_2+n_2^2B_{n,22}.
\)
The term $n_2^2B_{n,22}$ is the accumulated contribution of the common group shock across all pairs of observations in group two.

By contrast, the algorithmic fluctuation of the uniformly sampled estimator around the finite-sample target is governed by the residuals $\epsilon_n=Q_nu_n$. Since the group indicators $M_n$ are included in the design matrix, $Q_nM_n=0$. Hence, 
we have
\(
Q_n\Omega_nQ_n=Q_n.
\)
Thus the common group shock affects the sampling fluctuation of the full-sample estimator around $\theta^\star$, 
but it is not the residual fluctuation around $\theta_n$ that drives the uniform-sampling algorithmic law. In particular, for the group-two indicator,
\[
x^\top\operatorname{Diag}(\operatorname{diag}(Q_n\Omega_nQ_n))x
=
x^\top\operatorname{Diag}(\operatorname{diag}(Q_n))x
\le n_2,
\]
whereas $x^\top\Omega_nx=n_2+n_2^2B_{n,22}$. If $n_2/n\to\pi_2\in(0,1)$ and $B_{n,22}=\kappa/n$, then $n_2^2B_{n,22}\asymp n$, so the common group shock is first-order for inference on $\theta^\star$ even though it is invisible to the residual-based uniform-sampling fluctuation. This illustrates why
 the condition required for the validity of subsampling can fail.

\subsection{Proof of Proposition \ref{prop:unif-correlated-subrand}}
\label{sec:pf_unif-correlated-subrand}

\begin{proof}
By \Cref{prop:unif-correlated-clt},
\[
\Law\left(
\sqrt k(1-k/n)^{-1/2}\Sigma_n^{-1/2}(\widehat\theta_k-\theta_n)
\,\middle|\,
\mD_n
\right)
\Rightarrow_P
\mathcal N(0,I_d)
\]
for $k=m,b$. Since $m/n\to0$, $b/n\to0$, $\Sigma_n-G_n\to_P0$, and $G_n\to G\in\Spp$, it follows that
\[
\Law\left(
\sqrt k(\widehat\theta_k-\theta_n)
\,\middle|\,
\mD_n
\right)
\Rightarrow_P
\mathcal N(0,G),
\qquad k=m,b.
\]
Thus \Cref{ThconfintS} applies with $\tau_k=\sqrt k$. Finally, $\|\Omega_n\|=O(1)$ and $n^{-1}\mathbf X_n^\top \mathbf X_n\to\Psi\in\Spp$ imply $\theta_n-\theta^\star=O_P(n^{-1/2})$. Since $m/n\to0$,
\[
\sqrt m(\theta_n-\theta^\star)=o_P(1).
\]
The population-coverage claim follows from \Cref{ThconfintS}.

For the scalar contrast in part 2, apply \Cref{thm:politis-two-sided}
with deterministic studentizers $\widehat t_{m,n}=\sqrt m$ and
$\widehat t_{b,n,0}=\sqrt b$. For any fixed non-zero vector $c\in\mathbb R^d$, the conditional
limit above implies
\[
\Law\left(
\sqrt k\,c^\top(\widehat\theta_k-\theta_n)
\,\middle|\,
\mD_n
\right)
\Rightarrow_P
\mathcal N(0,c^\top Gc),
\qquad k=m,b.
\]
The relevant
Gaussian quantiles are regular. Since
$b<m$, we have $\sqrt m-\sqrt b>0$. Moreover, by
$\theta_n-\theta^\star=O_P(n^{-1/2})$ and $m/n\to0$,
\[
(\sqrt m-\sqrt b)c^\top(\theta_n-\theta^\star)
=O_P(\sqrt{m/n})=o_P(1).
\]
Thus \Cref{thm:politis-two-sided} gives nominal asymptotic
coverage for $c^\top\theta^\star$.

\end{proof}

{\small
\singlespacing
\setlength{\bibsep}{0.2pt plus 0.3ex}
\bibliographystyle{plainnat-abbrev}
\bibliography{references}
}

\end{document}